\documentclass{caosp309}

\usepackage{graphicx}
\usepackage{float}

\usepackage{natbib}
\articleNo{}
\pubyear{}
\volume{}
\volnumber{}
\firstpage{}
\received{}
\accepted{}

\def\BibTeX{{\rm B\kern-.05em{\sc i\kern-.025em b}\kern-.08em
             T\kern-.1667em\lower.7ex\hbox{E}\kern-.125emX}}

\begin{document}

%
\htitle{Al abundances of A-type main-sequence stars}
\hauthor{Y. Takeda}

\title{Photospheric aluminium abundances \\
of A-type main-sequence stars}


%
%
\author{
        Yoichi Takeda\orcid{0000-0002-7363-0447}
       }

%
\institute{
          11-2 Enomachi, Naka-ku, Hiroshima-shi, 730-0851, Japan \\
          \email{ytakeda@js2.so-net.ne.jp}
          }

\date{}

\maketitle

\begin{abstract}
Although anomalous surface abundances are often observed in A-type main-sequence 
stars (known as chemically peculiar stars; e.g., metallic line stars or Am stars), 
our understanding about the behavior of aluminium is still insufficient. Actually, 
even whether Al is overabundant or underabundant in Am stars is not clarified. 
This is presumably because most of the previous studies employed the Al~{\sc i} 3944/3961 
lines with the assumption of LTE, despite that a considerable non-LTE effect  
is expected in this resonance doublet. With an aim to shed light on this issue,
extensive statistical-equilibrium calculations on Al~{\sc i}/Al~{\sc ii} were carried 
out for a wide range of atmospheric parameters, based on which the non-LTE Al 
abundances were determined by applying the spectrum-fitting technique to the 
Al~{\sc i} 3944/3961 lines for 63 A-type dwarfs ($7000 \la T_{\rm eff} \la 10000$~K)  
of comparatively lower rotational velocities ($v_{\rm e}\sin i \la 100$~km~s$^{-1}$).
The following results were obtained.
(1) The non-LTE corrections ($\Delta$) are positive (reflecting the importance of 
overionization) and significantly large ($0.3 \la \Delta \la 1.0$~dex depending on 
$T_{\rm eff}$; generally $\Delta_{3944} < \Delta_{3961}$).
(2) By applying these corrections (and indispensable inclusion of Balmer 
line wings as background opacity), consistent non-LTE abundances for both lines 
could be obtained, and the serious zero-point discrepancy (considerably 
negative [Al/H] for normal metallicity stars of [Fe/H]~$\sim 0$) found in 
old studies has been settled.
(3) Al abundances of A-type stars are almost in proportion to [Fe/H] (tending to be 
overabundant in Am stars) with an approximate relation of [Al/H]~$\sim 1.2$~[Fe/H]. 
which is qualitatively consistent with the prediction of the diffusion 
theory (suggesting an Al excess in the photosphere of Am stars).
\end{abstract}

\section{Introduction}

An appreciable fraction of A-type stars ($7000 \la T_{\rm eff} \la 10000$~K) 
on the upper main sequence are chemical peculiar (CP) stars. Among these, 
A-type metallic-line (Am) stars are commonly observed in comparatively slower 
rotators ($v_{\rm e}\sin i \la 100$~km~$^{-1}$). While it is known that 
they show contrasting surface abundance anomalies between lighter and heavier
elements (e.g., deficiency in C, N, O, Ca, Sc; overabundances in Fe-group 
or s-process ones). abundance behaviors of those with intermediate atomic 
number ($10 < Z < 20$) are not necessarily well understood. 

One of such elements for which abundances are poorly determined is aluminium 
(Al, $Z = 13$). This is not due to the lack of available lines, as 
strong resonance doublet lines of neutral aluminium at 3944 and 3961~\AA\
are observable in the spectra of most A-type stars, although  
most Al atoms are in the once-ionized stage and only a tiny fraction
remains neutral in stellar atmospheres of this $T_{\rm eff}$ range (Fig.~1).
Actually, determinations of Al abundances in normal A-type and Am stars by 
using Al~{\sc i} 3944/3961 were reported already in the early work about 
a half century ago (e.g., Conti 1970; Smith 1971, 1973). 

\begin{figure}[H]
\centerline{\includegraphics[width=0.7\textwidth,clip=]{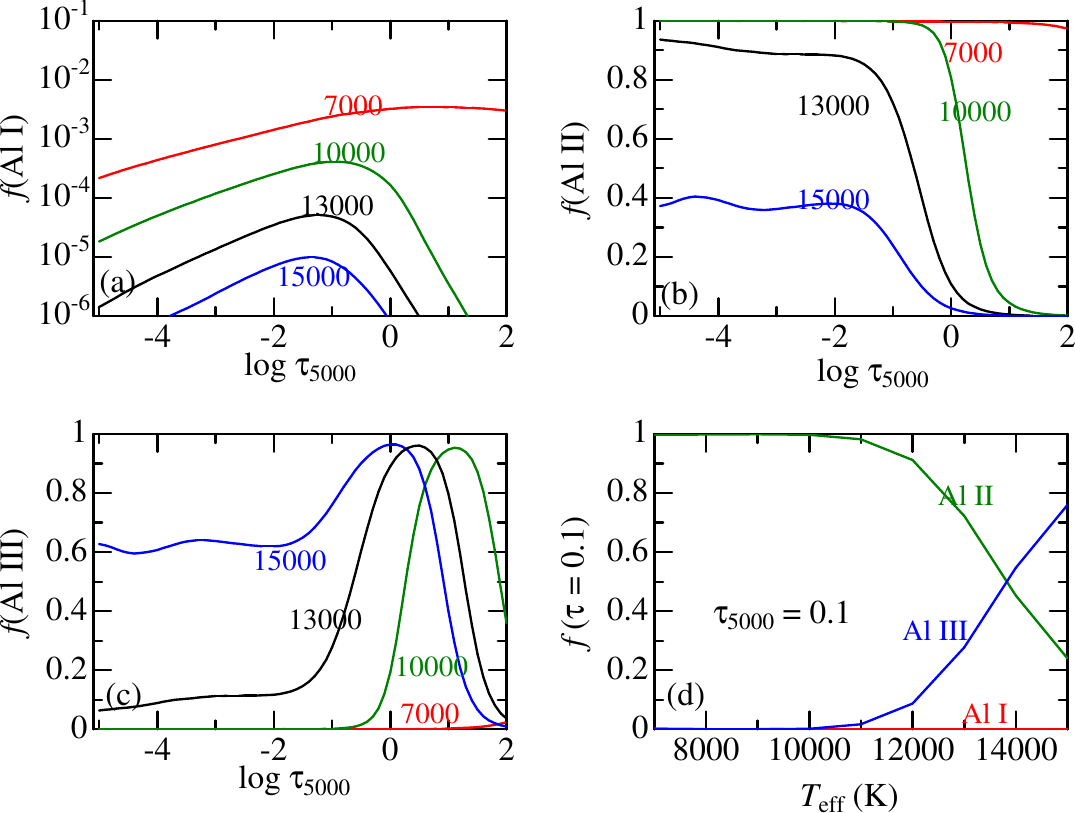}}
\caption{
Number population fraction ($f$) of (a) neutral, (b) once-ionized, 
and (c) twice-ionized aluminium species relative to the total Al atoms
[e.g., $f$(Al~{\sc i}) $\equiv N$(Al~{\sc i})/$N_{\rm total}^{\rm Al}$],
plotted against the continuum optical depth at 5000~\AA. 
Calculations were done for four $\log g = 4.0$ model of different
$T_{\rm eff}$ (7000, 10000, 13000, and 15000~K) as indicated in each panel.
The runs of $f$ for these three stages at $\tau_{5000} = 0.1$ with $T_{\rm eff}$ 
are also depicted in panel (d). All these calculations were done in LTE
(use of Saha's equation).
}
\label{fig1}
\end{figure}

\begin{figure}[H]
\centerline{\includegraphics[width=0.7\textwidth,clip=]{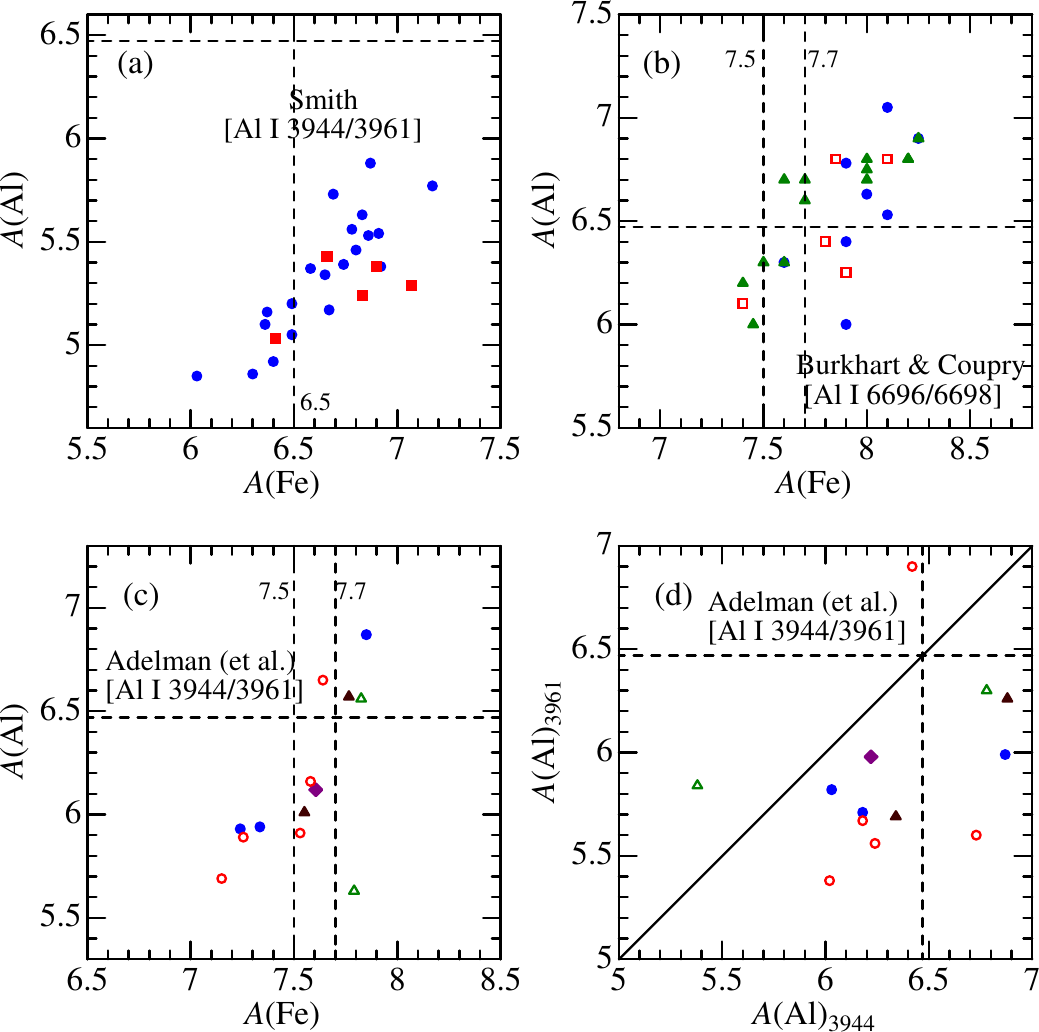}}
\caption{
Panels (a)--(c) show the correlations of Al and Fe abundances published 
in the past literature.
(a) Data of Smith (1971; circles) and Smith (1973; squares) based on 
Al~{\sc i} 3944/3961 lines. 
(b) Data of Burkhart and Coupry (1989--circles; 1991--triangles, 
2000--squares) based on Al~{\sc i} 6696/6698 lines.
(c) Data of Adelman (1984--filled diamond), Adelman et al. (1984--open triangles),
Adelman (1988--filled triangles), Adelman (1994--filled circles), 
and Adelman et al. (1997--open circles) based on Al~{\sc i} 3944/3961 lines.
(d) $A_{3944}$ vs. $A_{3961}$ relation for Adelman et al.'s data in panel (c).
In each panel, the locations of the solar Al abundance of $A_{\odot}$(Al) = 6.47 
(adopted in this study) as well as of the solar Fe abundances believed at the
time of the relevant papers (considerably revised over the past half century; 
cf. footnote~1) are indicated by dashed lines.
}
\label{fig2}
\end{figure}

Nevertheless, what was argued in those old studies was only qualitative that 
Al abundances in Am stars are comparatively higher than those in normal stars,
while nothing could be said about the quantitative extent of Al anomalies 
with respect to the reference normal (solar) abundance\footnote{
The solar photospheric abundance of aluminium $A_{\odot}$(Al) is considered 
to be well established ($A$ is the logarithmic number abundance of the element 
relative to that of hydrogen with the usual normalization of $A = 12.00$ for H), 
for which quite similar values have been reported so far: 6.40 (Lambert, Warner 1968), 
6.47 (Anders, Grevesse 1989), 6.45 (Asplund et al. 2009)  In this paper, 
Anders and Grevesse's $A_{\odot}$(Al) of 6.47 is adopted, as done in Kurucz's 
(1993) ATLAS9/WIDTH9 program. In contrast, $A_{\odot}$(Fe) has experienced 
considerable updates from low-scale to high-scale over the past half century, 
mainly due to the large revision in the transition probabilities of Fe lines 
(e.g., 6.5--7.0~$\rightarrow$~7.3--7.7~$\rightarrow \; \sim 7.5$; see Fig.~1 in 
Grevesse, Sauval 1999), though it is eventually settled around 7.50 
at present (cf. Asplund et al. 2009). 
} because absolute values of the resulting Al abundances were far from reliable. 

That is, since surface abundances of normal A-type stars of population I without 
chemical peculiarities should be more or less similar to the solar composition, 
[Al/H]~$\sim 0$ is expected to hold for stars of [Fe/H]~$\sim 0$.\footnote{
As usual, [X/H] is the differential abundance of element X relative to the Sun; 
i.e., [X/H]~$\equiv A_{\rm star}({\rm X})- A_{\odot}({\rm X})$.}
However, according to Fig.~ 2a, where the Al abundances of
A and Am stars derived by Smith (1971, 1973) from Al~{\sc i} 3944/3961 are 
plotted against the corresponding Fe abundances, the intercept of [Al/H]   
at [Fe/H]~$\sim 0$ is considerably subsolar ($\la -1$~dex), which suggests 
that his $A$(Al) values were significantly underestimated.  
This zero-point discrepancy ([Al/H]$<0$ at [Fe/H]~$\sim 0$) is similarly 
observed in Adelman et al.'s results (cf. Fig.~2c), who also employed 
the Al~{\sc i} 3944/3961 lines for deriving the Al abundances of A-type stars. 
To make things more complicated, the abundances they derived from these two  
lines are systematically discordant from each other ($A_{3961} < A_{3944}$;
cf. Fig.~2d). These are the problems involved with Al abundance 
determinations using these resonance lines.

Admittedly, this problem may be circumvented by invoking other Al lines.
As a matter of fact, Burkhart and Coupry (1989, 1991, 2000) employed the
high-excitation Al~{\sc i} 6696/6698 lines to derive Al abundances for 
late A-type and Am/Fm stars in the field and open clusters, and the resulting 
[Al/H] values appear to favorably satisfy the requirement ([Al/H]~$\sim 0$ 
at [Fe/H]~$\sim 0$) mentioned above (cf. Fig.~2b). Unfortunately, these lines are 
so weak that are usable only for sharp-lined lower $T_{\rm eff}$ ($\la 8000$~K) 
stars, which makes their applicability seriously limited. Accordingly, 
there is no other way than to avail of the strong Al~{\sc i} lines in the 
violet region if Al abundance behaviors of A-type stars in general are 
to be investigated.
   
Then, why are the Al abundances determined from 3944/3961 lines in the old 
studies unreliable and subject to appreciable errors (i.e., significantly 
underestimated)? The most likely reason is that they adopted the assumption of 
LTE in their analysis, because these Al~{\sc i} resonance lines are known to 
suffer a considerable non-LTE effect mainly caused by the overionization mechanism.
Actually, Steenbock and Holweger (1992) carried out a statistical-equilibrium
calculation in their analysis of Al~{\sc i} 3944 and 3961 lines of Vega and 
showed that rather large (positive) non-LTE corrections (+0.4 and +0.7~dex, 
respectively) should be applied. Since this correction acts in the direction 
of mitigating the underestimation, the problems involved in the Al abundances 
determined from 3944/3961 lines may be resolved by correctly taking into 
account the non-LTE effect.

However, ever since Steenbock and Holweger's (1992) non-LTE 
work confined to the specific case of Vega (mildly metal-deficient A0V star),
any investigation on the non-LTE effect of Al lines with regard to A-type stars 
in general has not been carried out to the author's knowledge, 
despite that not a few non-LTE studies on Al~{\sc i} line formation for late-type stars 
(FGK type stars of $T_{\rm eff} \la 6500$~K; especially in the metal-poor regime 
for studying galactic chemical evolution) have already been published in the 
past quarter century (Baum\"{u}ller, Gehren 1996, 1997; Mashonkina et al. 2008; 
Andrievsky et al. 2008; Nordlander, Lind 2017;  Ezzeddine et al. 2018). 

Motivated by this situation, the author decided to conduct extensive non-LTE
calculations on Al~{\sc i}+Al~{\sc ii} for a wide range of atmospheric parameters 
(covering main-sequence stars of mid-F through late-B types), and apply them 
to the analysis of Al~{\sc i} 3944/3961 lines for a sufficient number of 
A-type dwarfs by making use of the available high-dispersion spectra.
The points intended to clarify in this investigation are as follows.
\begin{itemize}
\item
How are the behaviors (e.g., characteristic trends, dependence upon atmospheric 
parameters) of the non-LTE effect for Al~{\sc i} 3944 ad 3961 resonance doublet 
lines in upper main-sequence stars? Is there any significant difference between 
these two lines? 
\item
To establish the Al abundance trends of A-type stars (Am and normal A stars) of various 
[Fe/H] by applying the spectrum-fitting analysis to the 3944/3961 
lines. Can the zero-point problem of [Al/H] seen in the old LTE work be removed 
by including the non-LTE corrections? Are the non-LTE abundances derived from 
3944 and 3961 lines consistent with each other? 
\item
Possibilities of Al abundance determination based on the other lines (than 
the Al~{\sc i} resonance lines of primary concern) are also examined,
such as Al~{\sc i} 6696/6698 lines (used by Burkhart and Coupry) or 
Al~{\sc ii} 3900/4663 lines (which have barely been employed in A-type stars). 
How are the non-LTE corrections for these lines?
\end{itemize}

\section{Non-LTE calculation for Al}

\subsection{Atomic model and computational details}

The statistical-equilibrium calculations for aluminium were carried out 
by using the non-LTE code described in Takeda (1991). 
The atomic model of Al adopted in this study was constructed 
based on Kurucz and Bell's (1995) compilation of atomic data ($gf$ values, 
levels, etc.), which consists of 65 Al~{\sc i} terms (up to 26$s$~$^{2}$S
at 48092~cm$^{-1}$) with 218 Al~{\sc i} radiative transitions, 
68 Al~{\sc ii} terms (up to 10$p$~$^{3}$P$^{\rm o}$
at 146602~cm$^{-1}$) with 601 Al~{\sc ii} radiative transitions, 
and 32 Al~{\sc iii} terms (up to 9$h$~$^{2}$H$^{\rm o}$ at 217252~cm$^{-1}$;
included only for conservation of total Al atoms). 
The ground $^{2}$P$^{\rm o}$ term of Al~{\sc i} (comprising two 
closely lying levels with statistical weights of $g$ = 2 and 4 correspondig to 
$J$ = 1/2 and 3/2) was treated as a single level with $g$ = 6. 

Regarding the calculation of photoionization rates, the cross-section data taken from 
TOPbase (Cunto, Mendoza 1992) were used for the lower 10 Al~{\sc i} terms and 
10 Al~{\sc ii} terms (while hydrogenic approximation was assumed for all other 
higher terms). 
As to the collisional (excitation and ionization) rates due to electron as well as
neutral hydrogen, the recipe described in Sect.~3.1.3 of Takeda (1991) was followed. 
Inelastic collisions due to neutral hydrogen atoms were formally included 
by the analytical formula as described therein 
with a moderate scaling factor of $S_{\rm H} = 0.4$ by following  Steenbock and 
Holweger (1992) as well as Baum\"{u}ller and Gehren (1997), though the effect 
of neutral hydrogen collision is insignificant in the atmosphere of early-type 
stars ($T_{\rm eff} \ga 7000$~K) under question. 

The calculations were done on a grid of 192 ($= 12 \times 4 \times 4$) Kurucz's 
(1993) ATLAS9 models atmospheres (scaled solar-abundance models according 
to the metallicity [Fe/H]) resulting from combinations of twelve $T_{\rm eff}$ 
values (6500, 7000, 7500, 8000, 8500, 9000, 9500, 10000, 11000, 12000, 13000, 
and 14000~K) and four  $\log g$ values (3.0, 3.5, 4.0, and 4.5), and
four [Fe/H] values ($-1.0$, $-0.5$, $0.0$, and $+0.5$)  while 
assuming $\xi$ = 2~km~s$^{-1}$ (microturbulence) and metallicity-scaled 
abundances of $A$(Al) = 6.47 + [Fe/H] (6.47 is the solar Al abundance used 
in Kurucz's ATLAS9 models) were adopted as the input Al abundance. 

\subsection{Behaviors of non-LTE departure coefficients}

The non-LTE departure coefficients of the representative four Al terms calculated 
for the model corresponding to Vega are plotted against the optical depth in Fig.~3. 
Comparing this figure with Fig.~2 of Steenbock and Holweger (1992), we can see 
that both calculations are reasonably consistent with each other.

\begin{figure}[H]
\centerline{\includegraphics[width=0.5\textwidth,clip=]{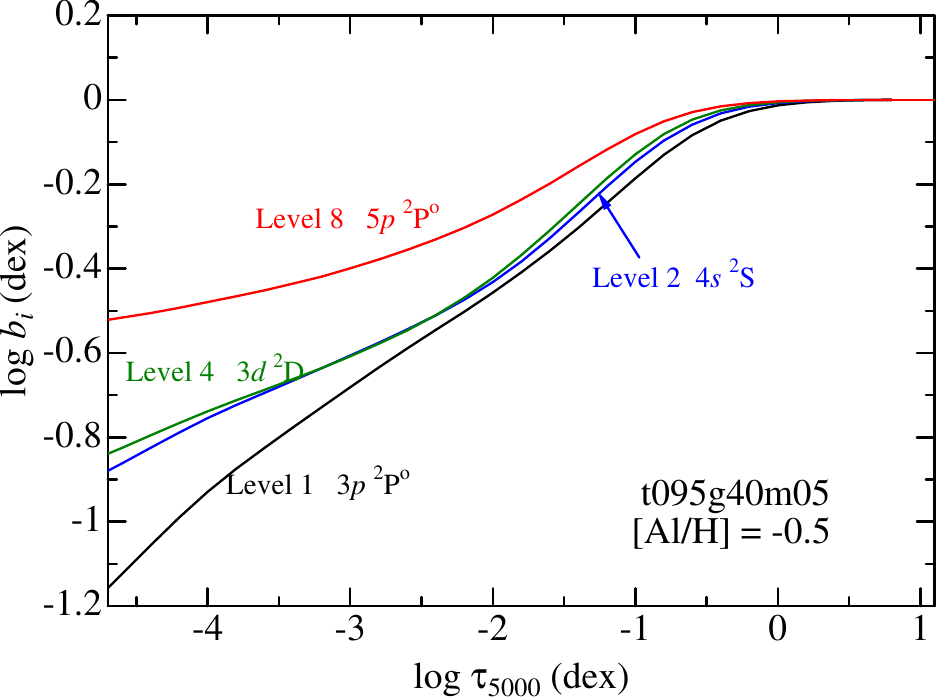}}
\caption{
Run of the non-LTE departure coefficients 
($b_{i} \equiv n_{i}^{\rm NLTE}/n_{i}^{\rm LTE}$) of neutral Al atom 
with the continuum optical depth at 5000~\AA\ calculated for
the model of $T_{\rm eff} = 9500$~K, $\log g = 4.0$, and [M/H] = [Fe/H] =
[Al/H] = $-0.5$.  Shown here are the results for the selected lower 4 terms
of $3p$~$^{2}$P$^{\rm o}$, $4s$~$^{2}$S, $3d$~$^{2}$D, and $5p$~$^{2}$P$^{\rm o}$
($i$ = 1, 2, 4, and 8). This figure is arranged so as to be compared with 
Steenbock and Holweger's (1992) Fig.~2.
}
\label{fig3}
\end{figure}

\begin{figure}[H]
\centerline{\includegraphics[width=0.8\textwidth,clip=]{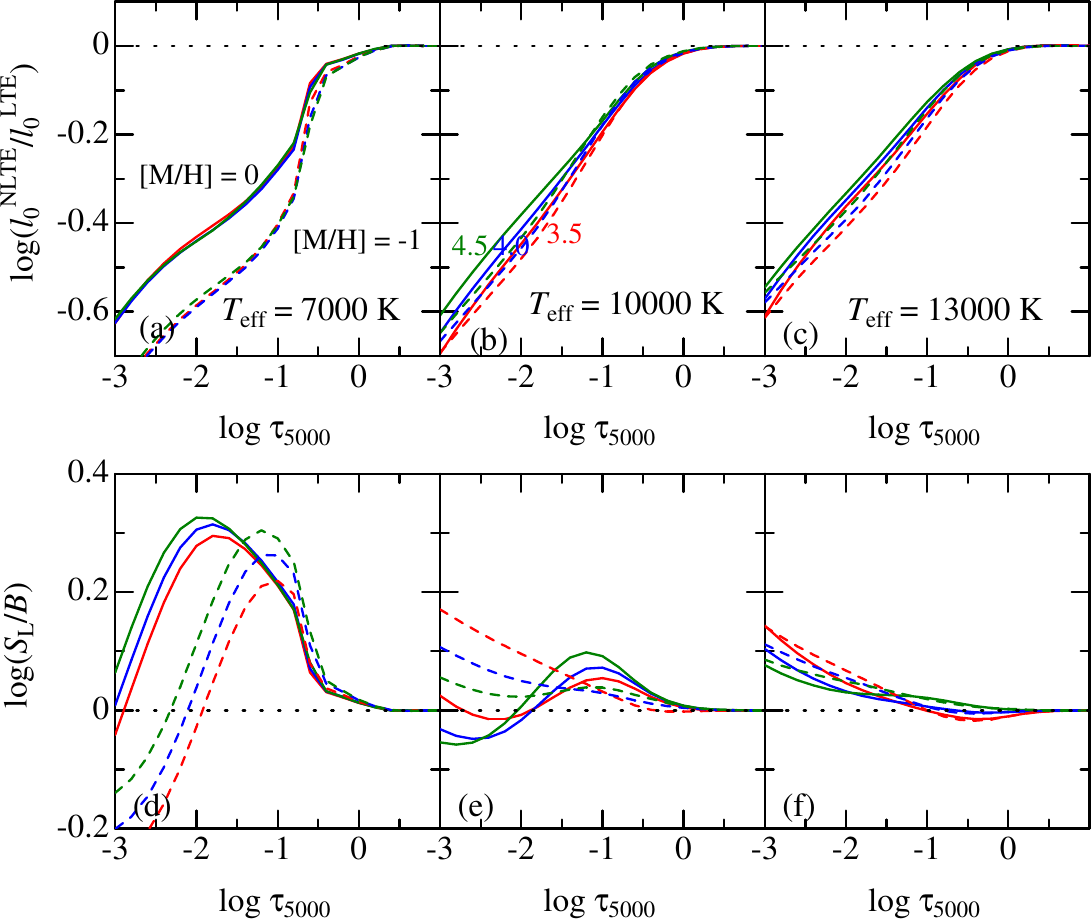}}
\caption{
The non-LTE-to-LTE line-center opacity ratio (upper panels a--c) and 
the ratio of the line source function ($S_{\rm L}$) 
to the local Planck function ($B$) (lower panels d--f)  
for the Al~{\sc i} $3p$~$^{2}$P$^{\rm o}$--$4s$~$^{2}$S transition 
(corresponding to Al~{\sc i} 3944/3961 lines) of multiplet~1, 
plotted against the continuum optical depth at 5000~\AA. 
Shown here are the calculations done with $\xi = 2$~km~s$^{-1}$ 
on the solar-metallicity models ([M/H] = [Fe/H] = 0; solid lines) and 
1/10$\times$ metal-deficient models ([M/H] = [Fe/H] = $-1$; dashed lines) 
of $T_{\rm eff} =$ 7000~K (left panels a, d), 10000~K (middle panels b, e), 
and 13000~K (right panels c, f), where metallicity-scaled Al abundance
([Al/Fe] = 0) was adopted in the calculation.
At each panel, the results for three $\log g$ values of 3.5, 4.0, and 4.5 
are depicted by different colors (red, blue, and green, respectively). 
}
\label{fig4}
\end{figure}

In Fig.~4 are also shown the $l_{0}^{\rm NLTE}(\tau)/l_{0}^{\rm LTE}(\tau)$ 
(the non-LTE-to-LTE line-center opacity ratio; almost equal to 
$\simeq b_{1}$) and $S_{\rm L}(\tau)/B(\tau)$ (the ratio of 
the line source function to the Planck function; nearly equal to 
$\simeq b_{2}/b_{1}$) for the transition relevant to the Al~{\sc i} 3944/3961 lines
($b_{1}$ and $b_{2}$ are the non-LTE departure coefficients for the lower and
upper terms), which were computed on the models of representative $T_{\rm eff}$ 
and $\log g$ values.  

As seen from this figure, the inequality relation $l_{0}^{\rm NLTE}/l_{0}^{\rm LTE} < 1$ 
(underpopulation or overionization) prevails in all depths, and the $S_{\rm L}/B \ga 1$  
(enhancement of line source function) tends to hold in the line-forming region,
both acting in the direction of weakening the strengths of absorption lines. 
Therefore, the Al~{\sc i} resonance doublet lines at 3944/3961~\AA\ are weakened 
by the non-LTE effect. Since this non-LTE underpopulation is due to the
overionization mechanism caused by the imbalance between the photoionization
and photorecombination rates ($J > B$), its extent may be sensitive to 
the metallicity ([M/H]) which generally plays an important role for the UV radiation field. 
According to Fig.~4, however, the effect of changing [M/H] is not so significant 
except for the case of lower $T_{\rm eff}$ (7000~K). 

\setcounter{table}{0}
\begin{table}[H]
\scriptsize
\caption{Adopted atomic data of Al lines.}
\begin{center}
\begin{tabular}
{cccccccc}\hline \hline
Species & Multiplet & $\lambda$ & $\chi_{\rm low}$ & $\log gf$ & Gammar & Gammas & Gammaw \\
  &  No.  &  (\AA)  & (eV)  &  (dex)  & (dex)  & (dex)  &  (dex) \\
\hline
Al~{\sc i} & 1  & 3944.006 & 0.000 & $-0.623$ & $(8.16)$ & $(-6.30)$ & $-7.32$ \\
Al~{\sc i} & 1  & 3961.520 & 0.014 & $-0.323$ & $(8.16)$ & $(-6.30)$ & $-7.32$ \\
Al~{\sc i} & 5  & 6696.023 & 3.143 & $-1.347$ & $(7.70)$ & $(-5.16)$ & $(-7.28)$ \\
Al~{\sc i} & 5  & 6698.673 & 3.143 & $-1.647$ & $(7.70)$ & $(-5.16)$ & $(-7.28)$ \\
\hline
Al~{\sc ii} & 1  & 3900.675 & 7.421 & $-1.270$ & 9.22 & $(-5.95)$ & $(-7.77)$ \\
Al~{\sc ii} & 2  & 4663.046 &10.598 & $-0.284$ & 7.99 & $(-5.53)$ & $(-7.64)$ \\
\hline
\end{tabular}
\end{center}
\scriptsize
Note. \\
These data are were taken from the VALD database (Ryabchikova et al. 2015),
while those parenthesized are the default values calculated
by Kurucz's (1993) WIDTH9 program.\\
Followed by first five self-explanatory columns,
damping parameters are given in the last three columns:\\
Gammar is the radiation damping width (s$^{-1}$), $\log\gamma_{\rm rad}$.\\
Gammas is the Stark damping width (s$^{-1}$) per electron density (cm$^{-3}$) 
at $10^{4}$ K, $\log(\gamma_{\rm e}/N_{\rm e})$.\\
Gammaw is the van der Waals damping width (s$^{-1}$) per hydrogen density 
(cm$^{-3}$) at $10^{4}$ K, $\log(\gamma_{\rm w}/N_{\rm H})$. \\
\end{table}

\subsection{Grid of abundance corrections}

Based on the results of these calculations, theoretical equivalent-widths 
and the corresponding non-LTE abundance corrections for Al~{\sc i} 3944 and 
3961 lines were computed for each of the models as follows.
First, for an assigned Al abundance ($A$ = 6.47 + [Fe/H]) and microturbulence 
($\xi$; any of 1, 2, 3, and 4~km~s$^{-1}$),\footnote{The departure 
coefficients computed for a fixed $\xi$ of 2~km~s$^{-1}$ were applied
also to the cases of $\xi$ = 1, 3, and 4~km~s$^{-1}$ because they are
not so sensitivity to a choice of $\xi$.} the non-LTE equivalent width 
($W^{\rm N}$) of the line was calculated by using the computed non-LTE departure 
coefficients ($b$) for each model atmosphere (LTE equivalent width $W^{\rm L}$ was 
also calculated for comparison, though not used for evaluation of $\Delta$). 
Next, the LTE ($A^{\rm L}$) and NLTE ($A^{\rm N}$) abundances were computed 
from this $W^{\rm N}$ while regarding it as if being a given observed 
equivalent width. We could then obtain the non-LTE abundance correction ($\Delta$), 
which is defined in terms of these two abundances as $\Delta \equiv A^{\rm N} - A^{\rm L}$.

Here, Kurucz's (1993) WIDTH9 program, which was considerably 
modified in many respects (e.g., incorporation of non-LTE departure 
in the line source function as well as in the line opacity, etc.), 
was employed for calculating the equivalent width for a given abundance, 
or inversely evaluating the abundance for an assigned equivalent width. 
The background opacities of overlapping Balmer lines, which are important
for these Al lines in the violet region (especially for the 3961 line), 
were included as done in Kurucz's (1993) ATLAS9 program (Griem 1960, 1967).
The adopted atomic data of these lines ($gf$ values, damping constants, etc.) 
are summarized in Table 1. 
The resulting grids of $W^{\rm L}$, $W^{\rm N}$, $A^{\rm N}$, $A^{\rm L}$, 
and $\Delta$ calculated for 3944 and 3961 lines are presented in the 
supplementary materials: 8 data files named as ``\verb|ncor####.%%%|'',
where `\verb|####|' denotes the relevant line (`3944' or `3961') and 
`\verb|%%%|' is the metallicity code (`m10', `m05', `p00', and `p05' corresponding 
to [X/H] = [Fe/H] = $-1.0$, $-0.5$, 0.0, and $+0.5$, respectively).

\begin{figure}[H]
\centerline{\includegraphics[width=0.7\textwidth,clip=]{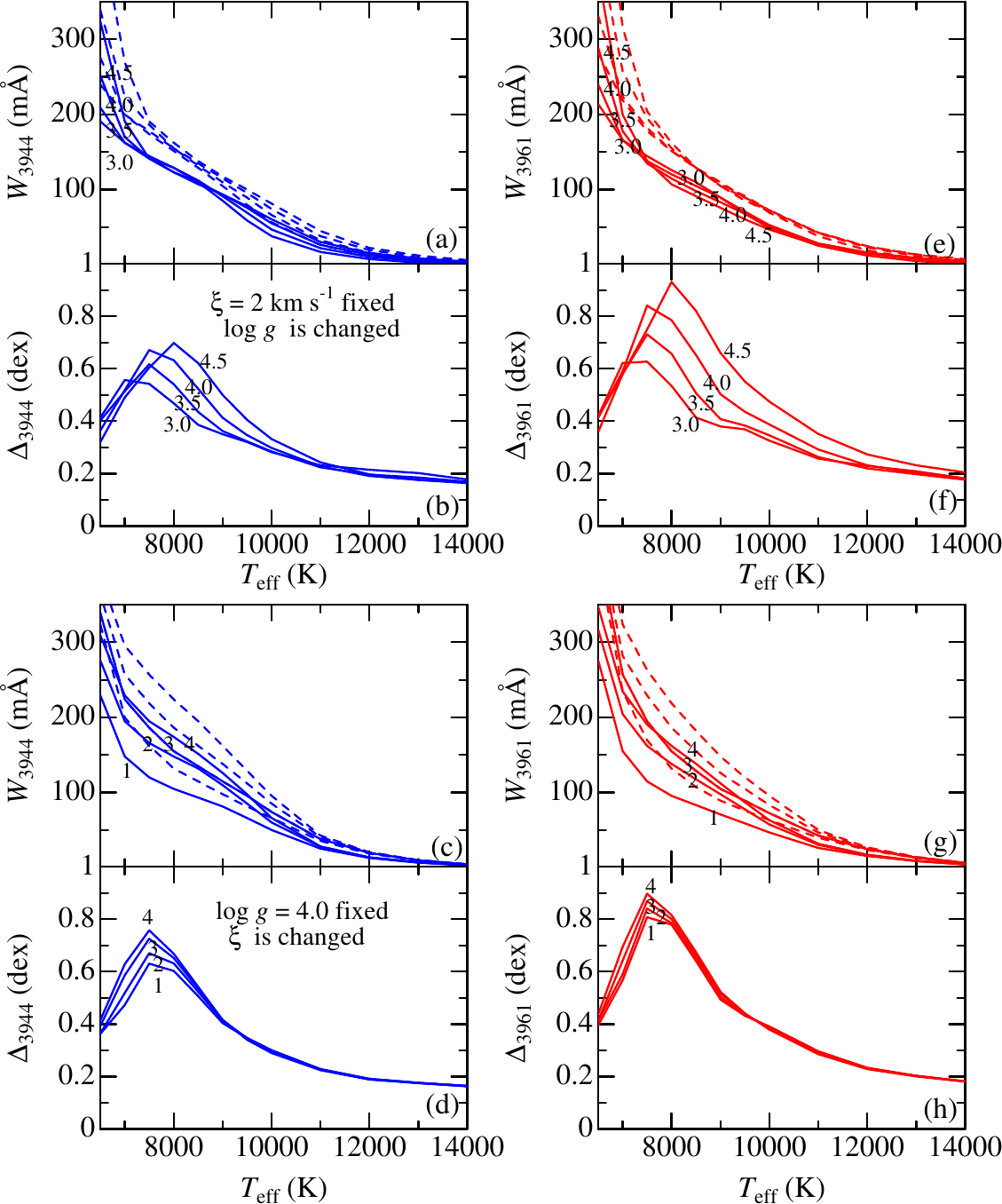}}
\caption{
The non-LTE and LTE equivalent widths ($W^{\rm N}$ and $W^{\rm L}$) 
for the Al~{\sc i} 3944/3961 lines and the corresponding non-LTE corrections 
($\Delta$), which were computed on the non-LTE grid of models described in 
Sect~2.1, are plotted against $T_{\rm eff}$.
Each figure set consists of two panels; the upper panel is for $W^{\rm N}$ 
(solid lines) and $W^{\rm L}$ (dashed lines), while the lower panel is
for $\Delta$. The upper sets (a+b, e+f) show the case of fixed $\xi$ 
(2~km~s$^{-1}$) but different $\log g$ (3.0, 3.5, 4.0, and 4.5), 
while the lower sets (c+d, g+h) are for the case of fixed $\log g$ (4.0) 
but different $\xi$ (1, 2, 3, and 4~km~s$^{-1}$). The left-hand figures 
show the results for the Al~{\sc i} 3944 line, while the right-hand
ones for the Al~{\sc i} 3961 line.
}
\label{fig5}
\end{figure}

\subsection{Dependence of $W$ and $\Delta$ upon stellar parameters}

How the theoretical $W$ and $\Delta$ computed for these two Al~{\sc i} lines 
depend upon the atmospheric parameters ($T_{\rm eff}$, $\log g$, and $\xi$) 
is illustrated in Fig.~5, from which the following characteristics are read.
\begin{itemize}
\item
The equivalent widths ($W$) progressively decrease with an increase 
in $T_{\rm eff}$, reflecting the $T$-dependence of 
$\propto \exp [\chi_{\rm ion}/(k T)]$ for the number population of 
the ground level ($k$ is the Boltzmann constant, $\chi_{\rm ion}$ is 
the ionization potential of 5.98~eV), while the inequality relation
of  $W^{\rm N} < W^{\rm L}$ (non-LTE line weakening) generally holds 
if compared at the same condition. Although the transition probability
($gf$ value) for the 3961 line is twice as stronger as that of the 3944
line (cf. Table~1), the difference between $W_{3944}$ and $W_{3961}$
is not so manifest, which is presumably because the latter $W_{3961}$ 
is more significantly affected by the opacity of H$\epsilon$ line 
wing (tending to weaken the line strength). 
\item
The non-LTE corrections are always positive ($\Delta > 0$) reflecting
that the line is weakened by the non-LTE effect, and significantly large 
($\sim$~0.2--1.0~dex). The $\Delta$ values appreciably depend upon $W$ 
(and thus upon $T_{\rm eff}$), tending to be larger with increasing $W$. 
This is because the line-forming region moves towards upper layer (where 
the departure from LTE is more significant) until desaturation begins 
in the flat-to-damping transition part of the curve of growth 
($W \ga 200$~m\AA\ at $T_{\rm eff} \la 8000$~K). 
Generally, $\Delta_{3944}$ is smaller than $\Delta_{3961}$, because
the latter 3961 line (larger opacity) forms in comparatively higher layer.
\item
$W$ tends to decline with a decrease in $\log g$, because ionization
is enhanced in the condition of lowered density. Since these lines
are generally strong ($W \ga 100$~m\AA) as to be in the flat part of 
the curve of growth, $W$ sensitively grows with an increase in $\xi$.
The reason why $\Delta$ tends to increase with $\log g$ as well as $\xi$
can be understood in terms of the $W$-dependence of $\Delta$ mentioned above. 
\end{itemize}

\section{Analysis of Al~{\sc i} 3944/3947 lines for A-type stars}

\subsection{Observational data}

Our next task is to determine the Al abundances for a number of sample 
stars from the Al~{\sc i} resonance doublet lines at 3944 and 3961~\AA\ 
by taking into account the non-LTE effect.
Regarding the observational data for this purpose, Takeda et al.'s 
(2008, 2009) spectra of A-type stars obtained by BOES (Bohyunsan Observatory
Echelle Spectrograph) were mainly used (Group B in Table~2), because 
they cover the relevant violet region thanks to their wide wavelength coverage,
where stars are limited to those of low-to-moderate projected rotational 
velocities ($v_{\rm e}\sin i \le 100$~km~s$^{-1}$) as done in Takeda et al. (2018) 
or Takeda (2022).

In addition, the spectra of 7 sharp-lined particularly bright A-type stars were 
secured by new observations specifically directed to the shorter wavelength 
region (Group A in Table~2), which were carried out on 2016 November 2--6 
at Okayama Astrophysical Observatory by using the 188~cm reflector along with 
HIDES (HIgh Dispersion Echelle Spectrograph) in the mode of blue cross disperser.
The data reduction was done in the standard manner by using IRAF,\footnote{
  IRAF is distributed by the National Optical Astronomy Observatories,
  which is operated by the Association of Universities for Research
  in Astronomy, Inc. under cooperative agreement with
  the National Science Foundation.} by which the spectra covering 
  3300--5600~\AA\ with a resolving power of $R\sim 100000$ were obtained. 
The finally resulting 63 program stars (7 from Group~A and 56 from Group~B) 
are listed in Table~3. 

It should be remarked here that a fraction of the Group~B spectra (9 stars) 
may be problematic in the sense that lines in the short-wavelength 
region appear unusually weaker than expected (designated as ``stars with 
weak broad Ca~{\sc ii} K line'' in Takeda 2020). Since different echelle orders 
are closely packed and the count level is considerably low in the blue-violet 
region of BOES spectra, this effect might be due to the stray light within 
the spectrograph. Accordingly, the results derived for those stars (marked 
with parentheses and asterisks in Table~3) should be viewed with caution.

\setcounter{table}{1}
\begin{table}[H]
\scriptsize
\caption{Basic information of the adopted observational data.}
\begin{center}
\begin{tabular}{cccccc}\hline\hline
Group & Instr. & Obs. Time & Resolution & Applied lines & Reference \\
\hline
A & HIDES & 2017 Oct & 100000  & 3944/3961/3900/4663 &  see Sect.~3.1 \\
B & BOES & 2008 Jan/Sep, 2009 Jan & 45000  & 3944/3961/6696/6698/4663 & Takeda et al. (2008, 2009) \\
C & HIDES & 2008 Oct & 100000 & 6696/6698/4663 & Takeda et al. (2012) \\
D & HIDES & 2006 May & 100000 & 4663 & Takeda et al. (2007) \\
\hline
\end{tabular} 
\end{center}
The group, to which the observational data adopted for each star belongs,  
is indicated in Table~3/tableE1.dat (3944/3961 lines; A or B), tableE2.dat 
(6696/6698 lines;  B or C), tableE3.dat (3900 line; A), and tableE4.dat
(4663 line; A or B or C or D).
\end{table}

\setcounter{table}{2}
\begin{table}[H]
\scriptsize
\caption{Program stars and their atmospheric parameters.}
\begin{center}
\begin{tabular}{cccccccccccc}\hline\hline
HD\# &  Name & Sp.Type & $T_{\rm eff}$ & $\log g$ & [Fe/H] & $\xi$ & $v_{\rm e}\sin i$ & 
Group & SB/V & CP & Hyades \\
(1) & (2) & (3) & (4) & (5) & (6) & (7) & (8) & (9) & (10) & (11) & (12) \\
\hline
   018454 &  4~Eri            & A5IV/V    &   7740& 4.07& +0.24 & 3.9& 100 & B & V    &   &     \\
   076543 &  $o^{1}$~Cnc      & A5III     &   8330& 4.18& +0.38 & 3.9&  91 & B & SB   &   &     \\
   012216 &  50~Cas           & A2V       &   9553& 3.90& +0.15 & 2.6&  88 & B & SB2  &   &     \\
   028355 &  79~Tau           & A7V       &   7809& 3.98& +0.19 & 4.0&  87 & B & V?   &   &  H  \\
   222345 &  $\omega^{1}$~Aqr & A7IV      &   7487& 3.88& $-$0.07 & 3.8&  86 & B & SB   &   &   \\
   074198 &  $\gamma$~Cnc     & A1IV      &   9381& 4.11& +0.25 & 2.8&  85 & B & SB   &   &     \\
   027934 &  $\kappa^{1}$~Tau & A7IV-V    &   8159& 3.84& +0.02 & 4.0&  83 & B & SB?  &   &  H  \\
   025490 &  $\nu$~Tau        & A1V       &   9077& 3.93& $-$0.05 & 3.2&  82 & B &      &   &   \\
   029388 &  90~Tau           & A6V       &   8194& 3.88& $-$0.01 & 4.0&  82 & B & SB1  &   &  H  \\
   079469 &  $\theta$~Hya     & B9.5V     &  10510& 4.20& $-$0.02 & 1.4&  82 & B & SB   &   &     \\
   028226 &                   & Am        &   7361& 4.01& +0.31 & 3.6&  81 & B & SB2  & Am&  H  \\
   207098 &  $\delta$~Cap     & A5mF2 (IV)&   7312& 4.06& +0.21 & 3.6&  81 & B & SBo  & Am&     \\
   033641 &  $\mu$~Aur        & A4m       &   7961& 4.21& +0.18 & 4.0&  79 & B & V    & Am&     \\
   216627 &  $\delta$~Aqr     & A3V       &   8587& 3.59& $-$0.25 & 3.7&  79 & B & V    &   &     \\
   012111 &  48~Cas           & A3IV      &   7910& 4.08& $-$0.23 & 4.0&  76 & B & SBo  &   &     \\
  (*023281)&                   & A5m       &   7761& 4.19& +0.05 & 4.0&  76 & B &      & Am&     \\
   192640 &  29~Cyg           & A2V       &   8845& 3.86& $-$1.41 & 3.5&  74 & B & V    & $\lambda$~Boo&     \\
   011636 &  $\beta$~Ari      & A5V...    &   8294& 4.12& +0.15 & 3.9&  73 & B & SBo  &   &     \\
   005448 &  $\mu$~And        & A5V       &   8147& 3.82& $-$0.14 & 4.0&  72 & B &      &   &     \\
   173880 &  111~Her          & A5III     &   8567& 4.27& +0.22 & 3.8&  72 & B & SB?  &   &     \\
   017093 &  38~Ari           & A7III-IV  &   7541& 3.95& $-$0.23 & 3.8&  69 & B & V    &   &     \\
   028319 &  $\theta^{2}$~Tau & A7III     &   7789& 3.68& $-$0.13 & 4.0&  68 & B & SB1o &   &  H  \\
   095382 &  59~Leo           & A5III     &   8017& 3.95& $-$0.09 & 4.0&  68 & B &      &   &     \\
  (*140436)&  $\gamma$~CrB      & A1Vs      &   9274& 3.89& $-$0.27 & 3.0&  68 & B &      &   &     \\
   020320 &  $\zeta$~Eri      & A5m       &   7505& 3.91& $-$0.12 & 3.8&  67 & B & SBo  & Am&     \\
   013161 &  $\beta$~Tri      & A5III     &   7957& 3.68& $-$0.32 & 4.0&  65 & B & SB2o &   &     \\
   027045 &  $\omega^{2}$~Tau & A3m       &   7552& 4.26& +0.36 & 3.8&  62 & B & SB   & Am&     \\
   200499 &  $\eta$~Cap       & A5V       &   8081& 3.95& $-$0.17 & 4.0&  62 & B & V    &   &     \\
  (*029499)&                   & A5m       &   7638& 4.08& +0.29 & 3.9&  61 & B & V    & Am&  H  \\
   116656 &  $\zeta$~UMa      & A2V       &   9317& 4.10& +0.28 & 2.9&  59 & B & SB2o &   &     \\
   198639 &  56~Cyg           & A4me...   &   7921& 4.09& +0.02 & 4.0&  59 & B & V?   & Am&     \\
  (*130841)&  $\alpha^{2}$~Lib & A3IV      &   8079& 3.96& $-$0.24 & 4.0&  58 & B & SB   &   &     \\
   030121 &  4~Cam            & A3m       &   7700& 3.98& +0.27 & 3.9&  57 & B &      & Am&     \\
   029479 &  $\sigma^{1}$~Tau & A4m       &   8406& 4.14& +0.35 & 3.9&  56 & B & SBo  & Am&  H  \\
   030210 &                   & Am...     &   7927& 3.94& +0.40 & 4.0&  56 & B & SB1? & Am&  H  \\
   222603 &  $\lambda$~Psc    & A7V       &   7757& 3.99& $-$0.17 & 4.0&  56 & B & SB   &   &     \\
   212061 &  $\gamma$~Aqr     & A0V       &  10384& 3.95& $-$0.08 & 1.5&  54 & B & SB   &   &     \\
   089021 &  $\lambda$~UMa    & A2IV      &   8861& 3.61& +0.08 & 3.5&  52 & B & V    &   &     \\
   195725 &  $\theta$~Cep     & A7III     &   7816& 3.74& +0.16 & 4.0&  49 & B & SB2o &   &     \\
   043378 &  2~Lyn            & A2Vs      &   9210& 4.09& $-$0.15 & 3.0&  46 & B & V?   &   &     \\
   027819 &  $\delta^{2}$~Tau & A7V       &   8047& 3.95& $-$0.05 & 4.0&  45 & B & SB   &   &  H  \\
   095418 &  $\beta$~UMa      & A1V       &   9489& 3.85& +0.24 & 2.7&  44 & B & SB   &   &     \\
  (*218396)&                   & A5V       &   7091& 4.06& $-$0.59 & 3.3&  41 & B &      &   &     \\
   084107 &  15~Leo           & A2IV      &   8665& 4.31& +0.01 & 3.7&  38 & B &      &   &     \\
   204188 &                   & A8m       &   7622& 4.21& +0.02 & 3.9&  36 & B & SBo  & Am&     \\
  (*033204)&                   & A5m       &   7530& 4.06& +0.18 & 3.8&  34 & B &      & Am&  H  \\
   141795 &  $\epsilon$~Ser   & A2m       &   8367& 4.24& +0.25 & 3.9&  32 & B & V    & Am&     \\
   173648 &  $\zeta^{1}$~Lyr  & Am        &   8004& 3.90& +0.32 & 4.0&  32 & B & SB1o & Am&     \\
  (*027628)&  60~Tau           & A3m       &   7218& 4.05& +0.10 & 3.5&  30 & B & SB1o & Am&  H  \\
   028546 &  81~Tau           & Am        &   7640& 4.17& +0.23 & 3.9&  28 & B & V?   & Am&  H  \\
   182564 &  $\pi$~Dra        & A2IIIs    &   9125& 3.80& +0.39 & 3.1&  27 & A &      &   &     \\
   172167 &  $\alpha$~Lyr     & A0Vvar    &   9435& 3.99& $-$0.53 & 2.7&  22 & A & V    & $\lambda$~Boo?&     \\
   060179 &  $\alpha$~Gem     & A2Vm      &   9122& 3.88& $-$0.02 & 3.2&  19 & B & SB1o & Am&     \\
   058142 &  21~Lyn           & A1V       &   9384& 3.74& $-$0.05 & 2.8&  19 & A & V    &   &     \\
   095608 &  60~Leo           & A1m       &   8972& 4.20& +0.31 & 3.3&  18 & B &      & Am&     \\
   048915 &  $\alpha$~CMa     & A0m...    &   9938& 4.31& +0.45 & 2.1&  17 & A & SBo  & Am&     \\
  (*027749)&  63~Tau           & A1m       &   7448& 4.21& +0.41 & 3.7&  13 & B & SB1o & Am&  H  \\
  (*033254)&  16~Ori           & A2m       &   7747& 4.14& +0.28 & 3.9&  13 & B & SBo  & Am&  H  \\
   072037 &  2~UMa            & A2m       &   7918& 4.16& +0.19 & 4.0&  12 & B &      & Am&     \\
   040932 &  $\mu$~Ori        & Am...     &   8005& 3.93& $-$0.12 & 4.0&  11 & B & SB1o & Am&  H  \\
   027962 &  $\delta^{3}$~Tau & A2IV      &   8923& 3.94& +0.25 & 3.4&  11 & A & SB   &   &  H  \\
   047105 &  $\gamma$~Gem     & A0IV      &   9115& 3.49& $-$0.03 & 3.2&  11 & A & SB   &   &     \\
   214994 &  $o$~Peg          & A1IV      &   9453& 3.64& +0.18 & 2.7&   6 & A & V    &   &     \\
\hline
\end{tabular} 
\end{center}
(1) HD number. Regarding those parenthesized with asterisks (9 stars), caution 
should be taken because their observational data may be unreliable (cf. Sect.~3.1). 
(2) Bayer/Flamsteed name. (3) Spectral type taken from Hipparcos catalogue (ESA 1997).
(4) Effective temperature (in K). (5) Logarithmic surface gravity ($\log g$ in dex,
where $g$ is in unit of cm~s$^{-2}$). (6) Differential Fe abundance relative to Procyon 
($\simeq$ Sun) derived as $A$(Fe) $- 7.49$.
(7) Microturbulent velocity (in km~s$^{-1}$).
(8) Projected rotational velocity (in km~s$^{-1}$). 
(9) Data source group used for the analysis of Al~{\sc i} 3944/3961 lines (cf. Table~2).
(10) Key to spectroscopic binary (SB, ``o'' denotes the case where orbital elements 
are available) or radial velocity variable (V). 
(11) Key to chemical peculiarity type (Am or $\lambda$~Boo), where the spectral 
classifications in three sources were consulted: Hipparcos catalogue (ESA 1997), 
Bright Star Catalogue (Hoffleit, Jaschek 1991), and SIMBAD.  
(12) Key to membership of Hyades cluster (H). Since these data (arranged in the descending
order of $v_{\rm e}\sin i$) are essentially the subset of those used in Takeda et al. (2018), 
the caption of Table~1 of that paper may be consulted for more details.
\end{table}

\subsection{Atmospheric parameters}

The atmospheric parameters of the program stars are presented in Table~3, 
which are the same as determined and adopted in Takeda et al. (2008, 2009).
Since only brief descriptions are given here, these papers should be consulted 
for more details:  
$T_{\rm eff}$ and $\log g$ are from Str\"{o}mgren's $uvby\beta$ color indices 
by using Napiwotzki et al.'s (1993) calibration, $\xi$ is from the empirical 
$T_{\rm eff}$-dependent relation [cf. Eq.(1) of Takeda et al. 2008], and [Fe/H] 
(Fe abundance, representative of the metallicity) is from the spectrum 
fitting analysis in the 6140--6170~\AA\ region.

The model atmosphere for each star was generated by interpolating
Kurucz's (1993) ATLAS9 grid of models in terms of $T_{\rm eff}$, $\log g$,
and [Fe/H]. Likewise, the depth-dependent non-LTE departure coefficients
to be used for each of the program stars were obtained by interpolation 
of the calculated grids (Sect. 2.1) with respect to $T_{\rm eff}$, 
$\log g$, and [Fe/H].

The 63 program stars are plotted on the $\log T_{\rm eff}$ vs. $\log L$ 
diagram in Fig.~6, where the theoretical evolutionary tracks of different masses
are also depicted for comparison. We can see from this figure that these 
stars are in the mass range of $1.5 M_{\odot} \la M \la 3 M_{\odot}$.

\begin{figure}[H]
\centerline{\includegraphics[width=0.5\textwidth,clip=]{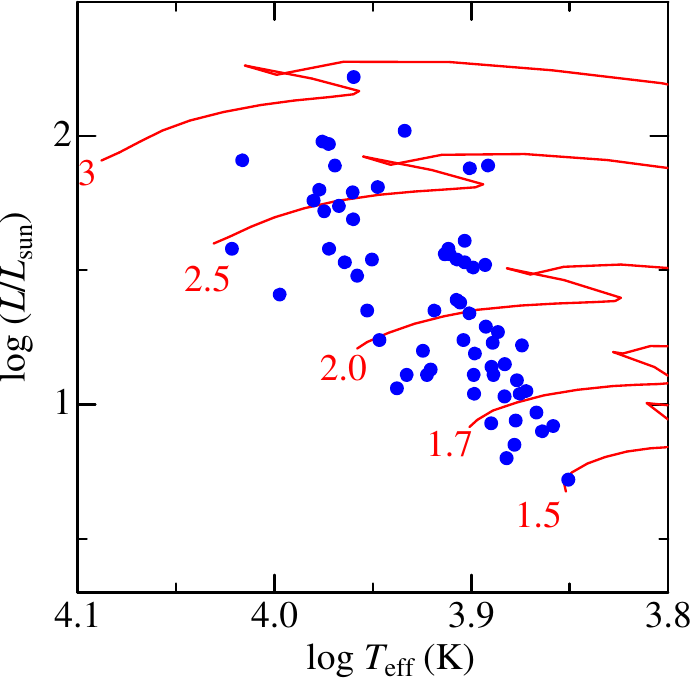}}
\caption{
Our program stars plotted on the $\log (L/L_{\odot})$ vs. $\log T_{\rm eff}$ 
diagram, where $L$ (luminosity) was evaluated from visual magnitude (corrected 
for interstellar extinction by following Arenou et al. 1992), Hipparcos parallax 
(van Leeuwen 2007), and bolometric correction (Flower 1996). Lejeune and Schaerer's 
(2001) theoretical solar-metallicity tracks for 5 different masses (1.5, 1.7, 2, 2.5, 
and 3~$M_{\odot}$) are also depicted by solid lines for comparison.
}
\label{fig6}
\end{figure}

\subsection{Abundance determination}

The abundances of Al for each of the 63 stars were determined in the similar 
manner as done in the previous papers (Takeda et al. 2008, 2009, 2018; Takeda 2022). 
(i) First, Takeda's (1995) spectrum-fitting technique is applied to each of the 
spectral regions comprising Al~{\sc i} 3944 and 3961 lines, while varying 
several free parameters ($v_{\rm e}\sin i$, radial velocity, and abundances 
of important elements; see Table~4 for more information), and the best fit 
parameter solutions are determined. (ii) Then, based on such established 
abundance solutions, the equivalent widths of these two Al lines ($W$) were 
inversely calculated. (iii) Finally, non-LTE ($A^{\rm N}$) and LTE ($A^{\rm L}$)
abundances were derived from $W$, along with the corresponding non-LTE correction 
$\Delta (\equiv A^{\rm N} - A^{\rm L})$.
The accomplished fit between the theoretical and observed spectra is displayed 
for each region in Fig.~7, and the results of the analysis ($W$, $A^{\rm N}$,
$A^{\rm L}$, $\Delta$) for each line are summarized in ``tableE1.dat'' of
the supplementary materials. 

\setcounter{table}{3}
\begin{table}[H]
\scriptsize
\caption{Details of spectrum fitting analyses.}
\begin{center}
\begin{tabular}{ccccc}\hline\hline
Lines & Fitting range (\AA) & Abundances varied$^{*}$ & Targets & Figure \\
\hline
Al~{\sc i} 3944  & 3941.5--3946.5 &  Al, Fe  & All 63 stars &  Fig.~7 \\
Al~{\sc i} 3961 & 3859--3964 &  Al, Fe  & All 63 stars & Fig.~7 \\
Al~{\sc i} 6696/6698   & 6692--6701 & Al, Fe & 8 sharp-lined late-A stars& Fig.~11a \\
Al~{\sc ii} 3900 & 3898.5--3902 & Al, Ti, V, Fe &  6 sharp-lined early-A stars & Fig.~11c \\
Al~{\sc ii} 4663 & 4662--4665 & Al, Fe & 10 sharp-lined early-A stars & Fig.~11e \\
\hline
\end{tabular}
\end{center}
$^{*}$ The abundances of other elements than these were fixed by assuming [X/H] = [Fe/H] in the fitting.\\ 
\end{table}

The resulting $W_{3944}$, $\Delta_{3944}$, and $A^{\rm N}_{3944}$ along with
the abundance sensitivities to typical ambiguities in atmospheric parameters 
[$\delta_{T\pm}$ (abundance changes for $T_{\rm eff}$ perturbations by $\pm 3$\%), 
$\delta_{g\pm}$ (abundance changes for $\log g$ perturbations by $\pm 0.1$~dex), 
and $\delta_{\xi\pm}$ (abundance changes for $\xi$ perturbations by $\pm 30$\%)] 
are plotted against $T_{\rm eff}$ in Fig.~8. (Although the results only for 
the Al~{\sc i} 3944 line are shown here, those for the 3961 line are quite similar.)
Fig.~8a and 8b confirm the trends of $W$ and $\Delta$ already described in  Sect.~2.4.
It can be seen also from Fig.~8d and 8f that Al abundances derived from these
strong resonance lines are quite sensitive to $T_{\rm eff}$ and especially to $\xi$,
as expected.

\begin{figure}[H]
\centerline{\includegraphics[width=0.9\textwidth,clip=]{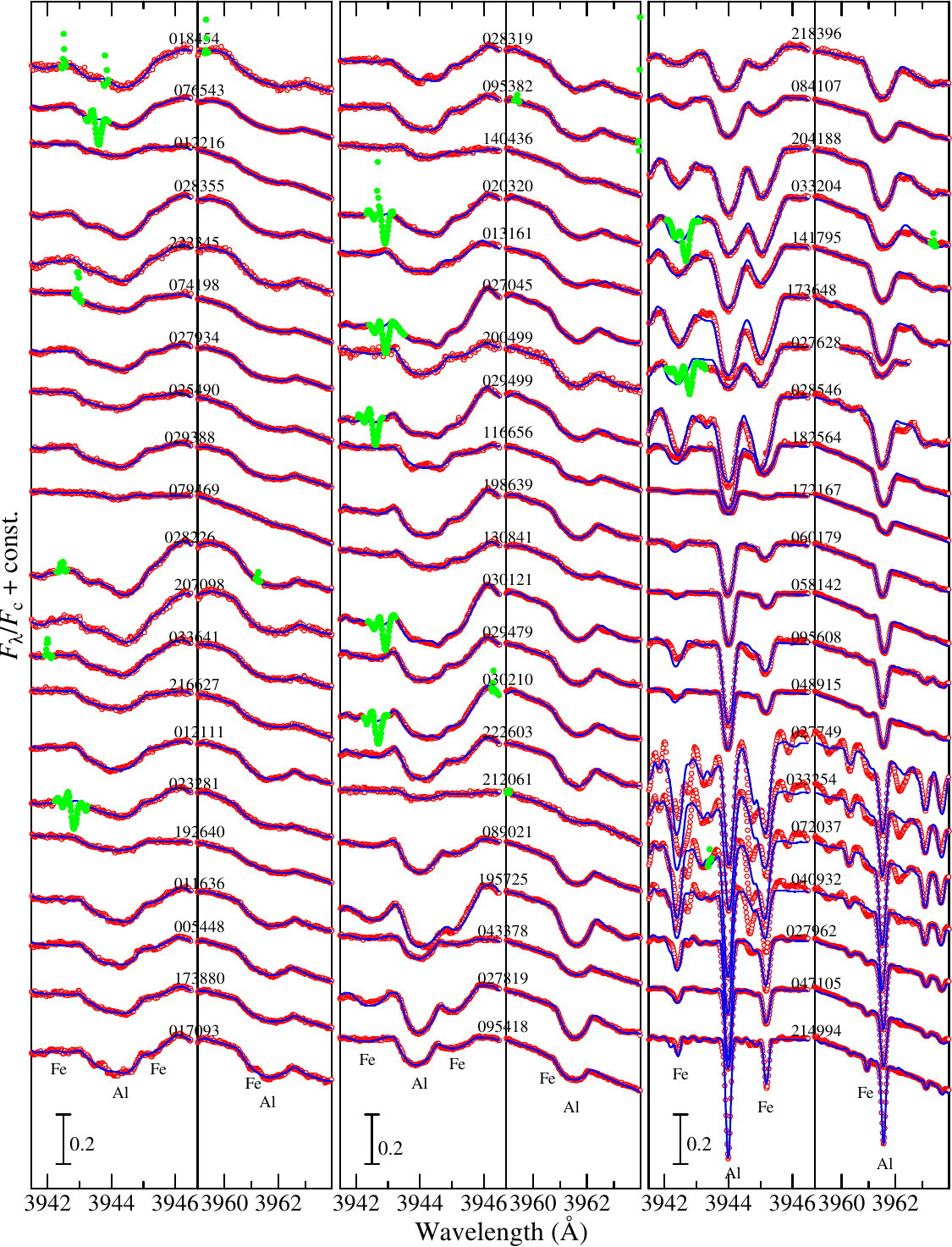}}
\caption{
Synthetic spectrum-fitting analysis for Al abundance determinations
from Al~{\sc i} 3944/3961 lines,
The best-fit theoretical spectra (in the selected ranges of 
3941.5--3946.5~\AA\ and 3959--3964~\AA\ comprising the relevant 
Al~{\sc i} lines) are depicted by blue solid lines, 
while the observed data are plotted by red symbols (the masked data 
excluded in judging the goodness of fit are highlighted in green).  
In each panel, the spectra (residual fluxes $F_{\lambda}/F_{\rm c}$;
the spectrum at 3946.5~\AA\ in the left and that at 3959~\AA\ in the 
right are so adjusted as to coincide with each other) are arranged 
in the descending order of $v_{\rm e} \sin i$ (from left to right, from top 
to bottom). An appropriate offset is applied to each spectrum (indicated 
by the HD number) relative to the adjacent one. 
The wavelength scale is in the laboratory frame after correcting 
the radial velocity shift.  
}
\label{fig7}
\end{figure}

\begin{figure}[H]
\centerline{\includegraphics[width=0.6\textwidth,clip=]{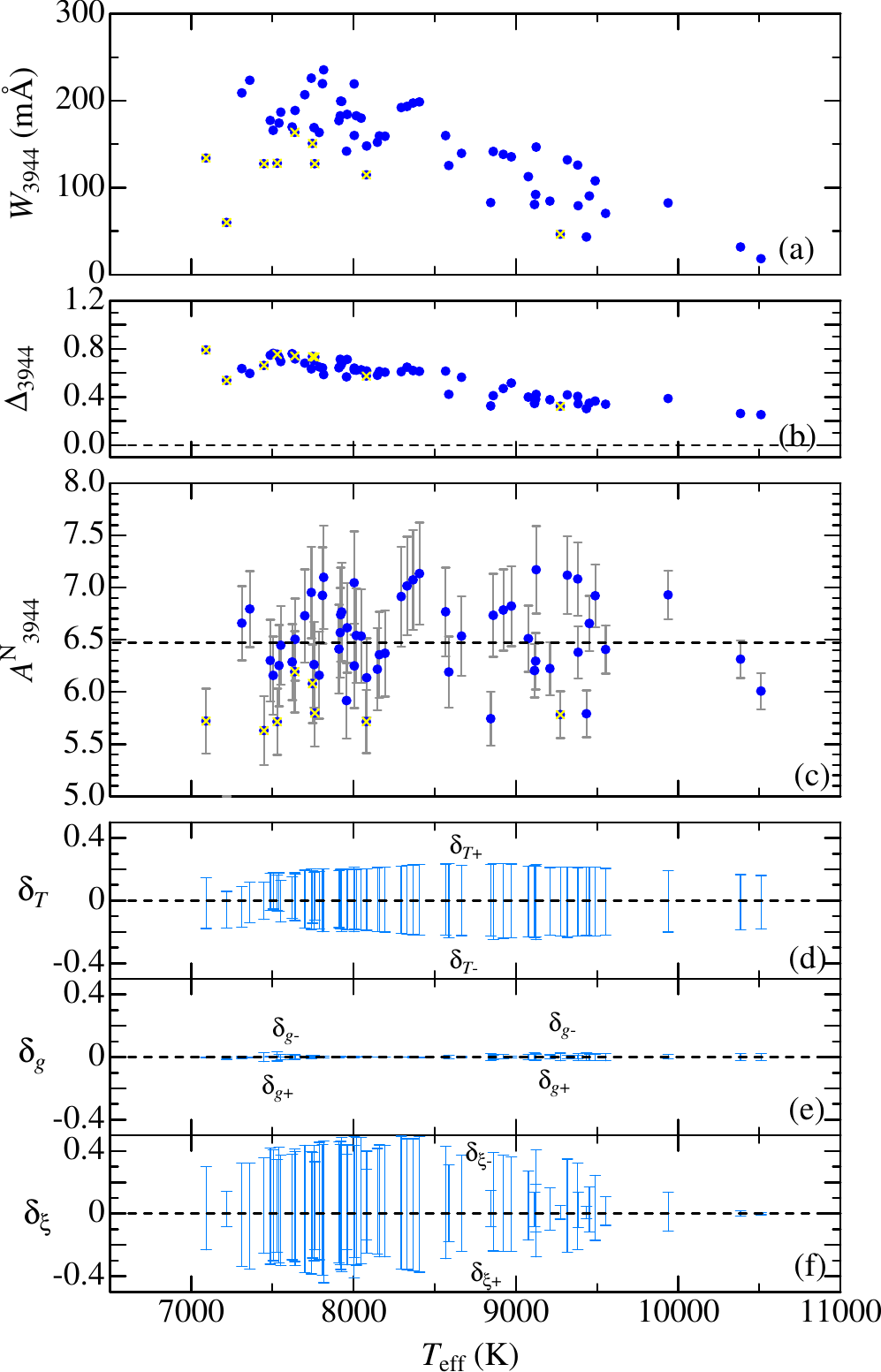}}
\caption{
Al abundances and the related quantities of the program stars, which were
derived from the Al~{\sc i} 3944 line, are plotted against $T_{\rm eff}$. 
(a) Equivalent widths ($W_{3944}$, filled symbols).
(b) Non-LTE corrections ($\Delta_{3944}$, filled symbols).
(c) $A^{\rm N}_{3944}$ (non-LTE Al abundances), where the adopted solar abundance 
($A_{\odot} = 6.47$) is indicated by the horizontal dashed line and the error 
bar denotes $\pm\delta_{Tgv}$ defined as the root-sum-square of $\delta_{T}$, 
$\delta_{g}$, and $\delta_{\xi}$ (e.g., $\delta_{T}$ is the mean of 
$|\delta_{T+}|$ and $|\delta_{T-}|$; etc.).
(d) $\delta_{T+}$ and $\delta_{T-}$ (abundance variations 
in response to $T_{\rm eff}$ changes of +3\% and $-$3\%). 
(e) $\delta_{g+}$ and $\delta_{g-}$ (abundance variations 
in response to $\log g$ changes by $+0.1$~dex and $-0.1$~dex). 
(f) $\delta_{\xi +}$ and $\delta_{\xi -}$ (abundance variations 
in response to perturbing $\xi$ by +30\% and $-$30\%).
The results based on unreliable observational data (cf. Sect.~3.1) are 
distinguished by overplotting yellow crosses in panels (a)--(c). 
}
\label{fig8}
\end{figure}

\begin{figure}[H]
\centerline{\includegraphics[width=0.8\textwidth,clip=]{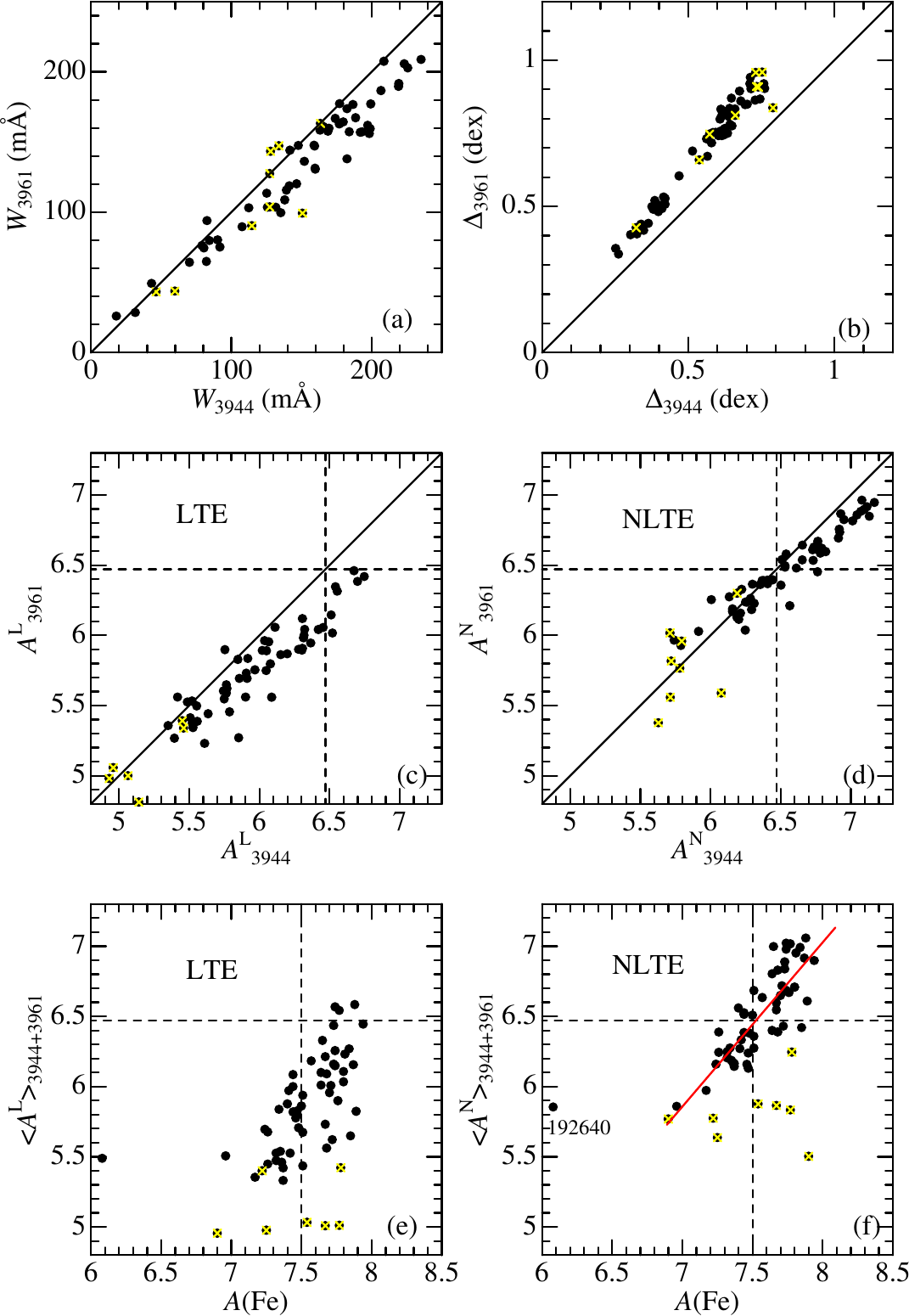}}
\caption{
Panels (a)--(d) show the correlations of the quantities between
Al~{\sc i} 3944 and 3961 lines (a: equivalent widths. b: NLTE corrections, 
c: LTE abundances, and d: NLTE abundances).
The mean Al abundances derived from Al~{\sc i} 3944/3961 lines,
defined as $\langle A \rangle_{3944+3961} \equiv (A_{3944} + A_{3961})/2$,
are plotted against $A$(Fe) (= [Fe/H] + 7.49; cf. Table~1) 
in panel (e; LTE) and panel (f; NLTE). 
The results based on unreliable observational data (cf. Sect.~3.1) are 
distinguished by overplotting yellow crosses as in Fig.~8.
In each panel, the locations of solar abundances are indicated
by dashed lines. The solid line in panels (a)--(d) is the guide line 
corresponding to $X_{3944} = X_{3961}$ ($X$ is $\Delta$ or $A$), 
while the red solid line depicted in panel (f) is the linear-regression 
line showing the main trend (cf. Sect.~3.4). 
}
\label{fig9}
\end{figure}

\subsection{Results and their characteristics}

The abundances and related quantities resulting from the analysis of the 
Al~{\sc i} 3944/3961 lines are graphically depicted in Fig.~9, from which 
the following trends are observed.
\begin{itemize}
\item
While the equivalent widths of these doublet lines($W_{3944}$ and $W_{3961}$) are not
much different from each other (Fig.~9a), the non-LTE corrections generally
satisfy the inequality relation $(0 <) \Delta_{3944} < \Delta_{3961}$ (Fig.~9b), 
as already mentioned in Sect.~2.4.
\item
Although $A^{\rm L}_{3944}$ tends to be systematically higher than  $A^{\rm L}_{3961}$
in the LTE case (Fig.~9c), this discrepancy is mitigated by the difference in $\Delta$ 
to result in $A^{\rm N}_{3944} \approx A^{\rm N}_{3961}$ in the non-LTE case (Fig.~9d). 
\item
In terms of the averaged abundance between two lines, 
$\langle A \rangle _{3944+3961} \equiv (A_{3944} + A_{3961}) / 2$,
while the LTE abundances ($\langle A^{\rm L} \rangle _{3944+3961}$) are considerably 
subsolar around the solar metallicity (Fig.~9e), this discordance is satisfactorily 
removed in the non-LTE case ($\langle A^{\rm N} \rangle _{3944+3961}$), so that  
normal-metallicity stars reasonably show $A_{\rm star} \sim A_{\odot}$ for both 
Al and Fe (Fig.~9f).  
\item 
By applying the least-squares analysis to the data in Fig.~9f (while excluding 
those yellow-crossed unreliable ones and that of the $\lambda$~Boo star HD~192640
showing exceptionally low metallicity of [Fe/H]= $-1.41$), the linear-regression 
relation [Al/H] $= 1.17 (\pm 0.11)$~[Fe/H]~$-0.04 (\pm 0.03)$ is obtained,
where [Al/H] = $A$(Al)~$-6.47$ and [Fe/H] = $A$(Fe)~$-7.49$.  
\item
The scaling relation ([Al/H] $\sim$ 1.2 [Fe/H]) observationally established here 
indicates that an overabundance of Al is associated with an increased Fe abundance in 
the surface of A-type stars. This trend is qualitatively consistent with the prediction 
from the diffusion theory, which suggests an abundance excess of Al (like Fe) 
as a result of element segregation process in the envelope of Am stars 
(Richer et al. 2000; Talon et al. 2006).   
\end{itemize}  

\subsection{Interpretation of the problems in previous studies}

It is now possible to discuss the problematic issues seen in the old Al abundance 
determinations for A-type stars based on the Al~{\sc i} resonance doublet lines.

The serious zero-point discrepancy in [Al/H] (considerably negative [Al/H] values 
for [Fe/H]~$\sim 0$ stars) seen in the Al abundances derived from Al~{\sc i} 3944/3961 
lines by Smith (1971, 1973) and Adelman (et al.) in 1980s--90s (cf. Figs.~2a and 2c)
is simply because they adopted LTE in their analysis. Since appreciable (positive) 
non-LTE corrections due to Al~{\sc i} overionization are required, LTE Al abundances 
are significantly underestimated and shifted towards subsolar direction (Fig.~9e). 
This problem can be reasonably resolved by taking into account the non-LTE effect 
as shown in Fig.~9f. 
  
The systematic discrepancy in the LTE abundances between Al~{\sc i} 3944 and 3961 
lines seen in Adelman et al.'s results ($A_{3944} > A_{3961}$; cf. Fig.~2d) may 
also be associated by the non-LTE effect, because neglecting non-LTE corrections 
($\Delta_{3944} < \Delta_{3961}$) results in such a tendency as shown in Fig.~9c.  
However, the extent of discordance observed in Fig.~2d seems to be larger than 
that expected from Fig.~9c. As another possibility, they might have
not included the opacity of Balmer line wings, the neglect of which can cause
appreciable abundance differences because the 3961 line is more affected by 
this effect (due to H$\epsilon$ at 3970~\AA) than the 3944 line, as demonstrated 
in Fig.~10. Therefore, in order to derive correct Al abundances from the 
Al~{\sc i} 3944/3961 lines, it is mandatory to take into account not only 
the non-LTE effect but also the background Balmer line opacities.  
   
\begin{figure}[H]
\centerline{\includegraphics[width=0.7\textwidth,clip=]{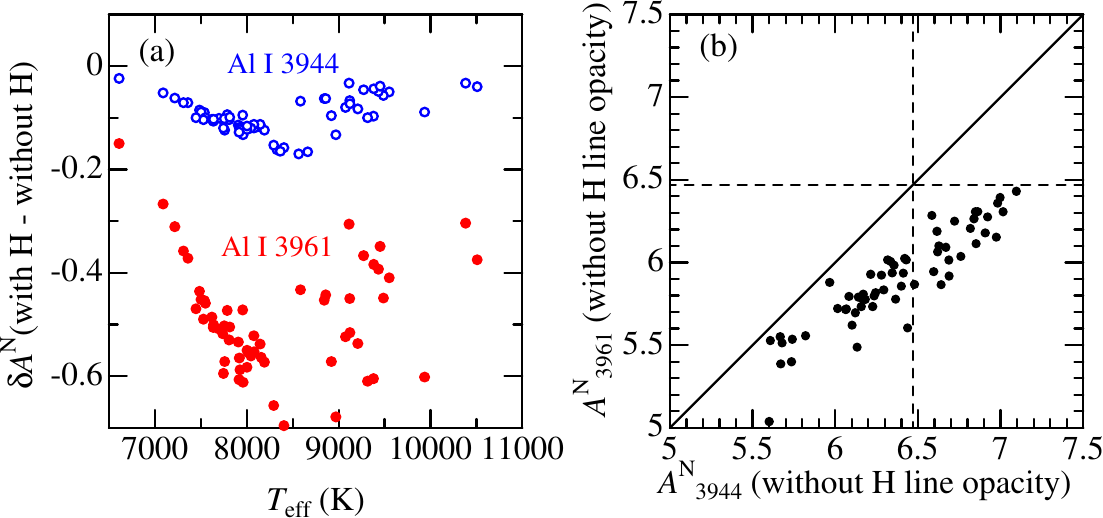}}
\caption{
(a) The values of $\delta A^{\rm N}$ (abundance difference of two  
$A^{\rm N}$ values obtained by neglecting and correctly 
including the overlapping opacity of Balmer line wings) are plotted
against $T_{\rm eff}$. Open and filed symbols correspond to Al~{\sc i} 3944
and 3961 lines, respectively.
(b) Correlation plots between $A_{3944}^{\rm N}$(without H line)
and $A_{3961}^{\rm N}$(without H line), which should be compared with 
Fig.~9d (results derived by correctly including the H line opacity).
}
\label{fig10}
\end{figure}

\section{Other aluminium lines as abundance indicators}

Finally, some discussion may be in order regarding whether and how 
other lines (than Al~{\sc i} 3944/3961 doublet) are applicable to determining 
Al abundances of A-type stars. Here, three candidate lines were examined:
Al~{\sc i} 6696/6698, Al~{\sc ii} 3900, and Al~{\sc ii} 4663.  
It turned out, however, that abundances could be derived for only a limited number 
($\la 10$ for each line) of sharp-line stars ($v_{\rm e}\sin i$ less than several 
tens km~s$^{-1}$) because these lines are generally weak. 

Accordingly, abundance determinations (such as done in Sect.~3.3) were carried out 
for 8 mostly late A-type stars (Al~{\sc i} 6696/6698), 6 early A-type stars 
(Al~{\sc ii} 3900), and 10 early A-type stars (Al~{\sc ii} 4663).
The adopted line data are given in Table~1, and the details of spectrum fittings 
are presented in Table~4. The results ($W$, $A^{\rm N}$, $A^{\rm L}$,
$\Delta$) are summarized in ``tableE2.dat'' (6696),\footnote{Although the fitting was 
done in the wavelength region including both 6696 and 6698 lines, only the results 
for the 6696 line (twice as strong as the 6698 line) are presented here.} 
``tableE3.dat'' (3900), and ``tableE3.dat'' (4663) of the supplementary materials.
The accomplished fit for each region and the correlations of the resulting non-LTE Al 
abundances ($A^{\rm N}_{6696}$, $A^{\rm N}_{3900}$, and $A^{\rm N}_{4663}$) with 
$\langle A^{\rm N} \rangle _{3944+3961}$ are displayed in Fig.~11.

Regarding the high-excitation Al~{\sc i} 6696 line, which was also used by Burkhart and 
Coupry  (1989, 1991, 2000) for their study of lower $T_{\rm eff}$ ($\la 8000$~K) stars,
$A^{\rm N}_{6696}$ is more or less consistent with $\langle A^{\rm N} \rangle _{3944+3961}$
(Fig.~11b), if unreliable (yellow-crossed) data and that of HD~047105 (the very weak-line case 
with $W_{6696}$ of only 1.2~m\AA) are excluded. It should be noted that positive non-LTE 
corrections of $\Delta \sim$~0.1--0.3~dex are expected for this line, though comparatively
less significant than the case of Al~{\sc i} 3944/3961. 
 
Rather disappointingly, the Al~{\sc ii} 3900.675 line was found to be badly blended with
the neighboring much stronger Ti~{\sc ii} 3900.539 line, as shown in Fig.~11c.
While Al abundances could be somehow determined in this synthetic fitting analysis
by making use of the slight asymmetry in the blended feature, the resulting $A^{\rm N}_{3900}$
turned out to be systematically lower than $\langle A^{\rm N} \rangle _{3944+3961}$ (Fig.~12d); 
the reason for this discrepancy is not clear. In any event, this Al~{\sc ii} 3900 line 
is not suitable for abundance determination, despite that it is unaffected by
any non-LTE effect ($|\Delta| \la 0.01$~dex, its sign is positive or negative).

In contrast, the other Al~{\sc ii} line at 4663~\AA\ is much more favorable, which is
almost free from any blending (Fig.~11e). Actually, $A^{\rm N}_{4663}$ and
$\langle A^{\rm N} \rangle _{3944+3961}$ are in satisfactory agreement with each other
(though one exception is HD~060179) as seen in Fig.~11f. Since the non-LTE correction
is negligibly small ($\Delta \la 0.02$~dex; $\Delta$ is positive), it may be analyzed
with the assumption of LTE. This Al~{\sc ii} 4663 line would serve as a good
Al abundance indicator for early A- through late B-type stars, as long as it is
measurable. 

\begin{figure}[H]
\centerline{\includegraphics[width=0.8\textwidth,clip=]{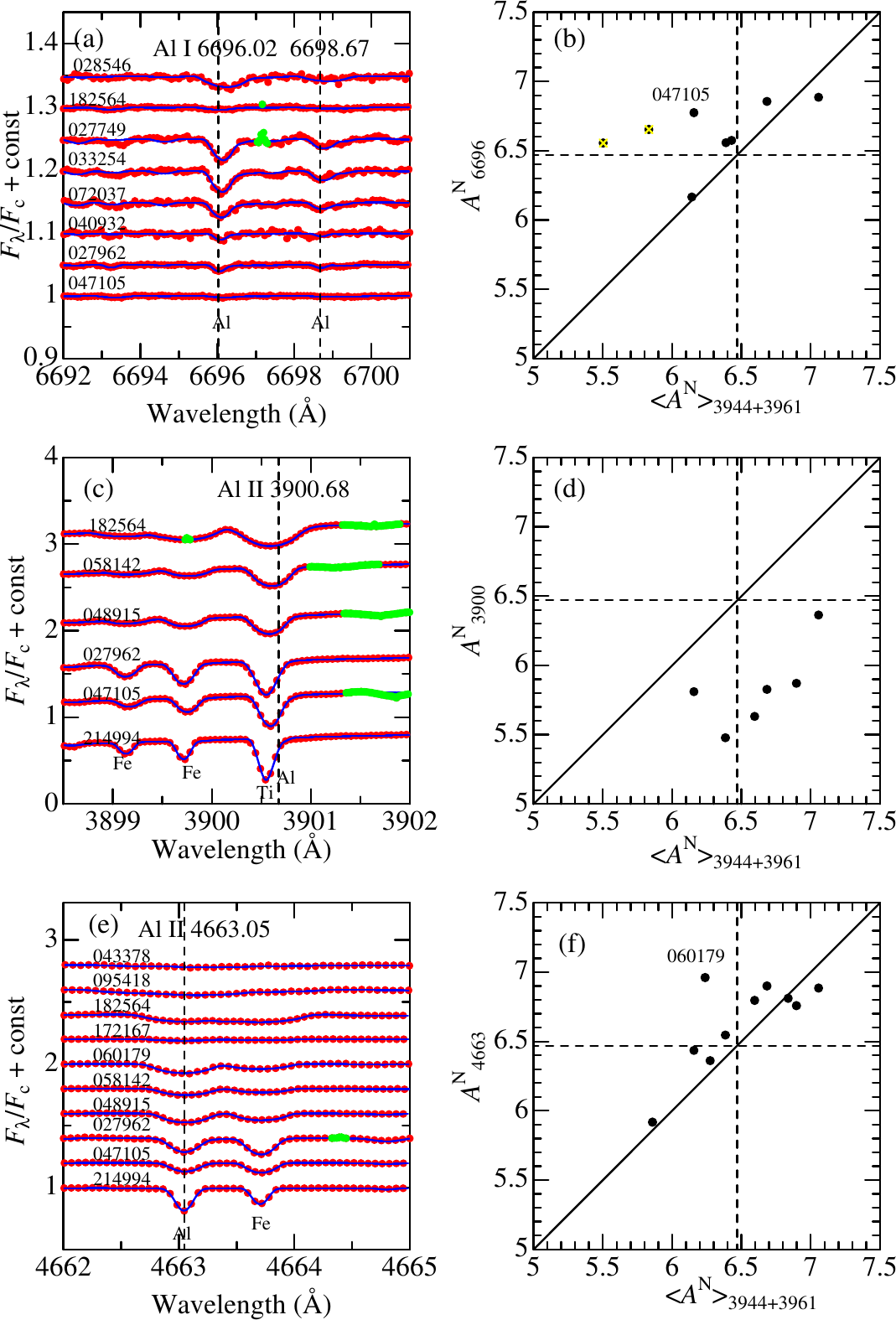}}
\caption{
The left-hand panels (a, c, e) show the accomplished fit of synthetic spectrum 
analysis carried out for determining the abundances of Al by using 
(a) Al~{\sc i} 6696/6698 (6692--6701~\AA\ region)
(c) Al~{\sc ii} 3900 (3898.5--3902~\AA\ region), and 
(e) Al~{\sc ii} 4663 (4662--4665~\AA\ region). 
The wavelength positions of the relevant Al lines are shown by vertical dashed lines.
Otherwise, the meanings of the lines and symbols are the same as in Fig.~7.
The non-LTE Al abundances ($A^{\rm N}$) resulting from these three regions 
are plotted against $\langle A^{\rm N} \rangle_{3944+3961}$ (mean NLTE abundances 
of those derived from Al~{\sc i} 3944/3961 lines) in the corresponding 
right-hand panels (b, d, f), respectively.  
Yellow crosses are overplotted on the results based on unreliable 
observational data as in Fig.~9.
}
\label{fig11}
\end{figure}

\section{Summary and conclusion}

Although various abundance anomalies are known to be observed in the surface 
of A-type stars on the upper main sequence (e.g., Am stars), the behaviors of 
aluminium abundances are still poorly understood, for which even the qualitative 
trend (excess or deficiency in Am stars) is not established. 

That is, according to the past work on Al abundances of A stars, for which 
the strong resonance Al~{\sc i} 3944/3961 lines were mostly employed, [Al/H] tends 
to be considerably negative for stars of near-normal metallicity ([Fe/H]~$\sim 0$),
despite that $A$(Al) appears to be positively correlated with $A$(Fe). 
Moreover, the reported Al abundances derived from 3944 and 3961 lines were not 
consistent with each other.

This is presumably related to the fact that these previous Al abundance determinations 
were done with the assumption of LTE, despite that considerable non-LTE corrections 
are suspected for these Al~{\sc i} lines. Unfortunately, however, the impact of 
non-LTE effect on Al abundance determination of A-type stars in general has barely 
been investigated so far.  

With an aim to shed light on this issue, extensive statistical-equilibrium calculations 
on Al~{\sc i}/Al~{\sc ii} were carried out for a wide range of atmospheric parameters
covering early F- to late B-type main-sequence stars (6500--14000~K in $T_{\rm eff}$, 
3.0--4.5 in $\log g$, and $-1.0$ to $+0.5$ in [Fe/H]), and the equivalent widths 
($W$) and non-LTE abundance corrections ($\Delta$) for these 3944 and 3961 lines 
were then calculated for these model grids, which are presented as supplementary 
materials.

These calculations revealed that these Al~{\sc i} resonance doublet lines are 
considerably weakened by the non-LTE effect due to the overionization mechanism, 
which means that $\Delta$ values are always positive and significantly large 
($0.3 \la \Delta \la 1.0$~dex) and tending to increase with $W$ (or with a decrease 
in $T_{\rm eff}$), where an inequality relation of $\Delta_{3944} < \Delta_{3961}$ 
generally holds.

As a practical application of these results, non-LTE Al abundances were 
determined by applying the spectrum-fitting technique to the Al~{\sc i} 3944/3961 
lines for selected 63 A-type dwarfs ($7000 \la T_{\rm eff} \la 10000$~K) of 
comparatively lower rotational velocities ($v_{\rm e}\sin i \la 100$~km~s$^{-1}$)  
based on the high-dispersion spectra obtained at Okayama Astrophysical Observatory 
and Bohyunsan Optical Astronomy Observatory.

It then turned out that consistent non-LTE abundances for both lines could be 
obtained and resulted in [Al/H] $\sim$ [Fe/H] $\sim 0$ for normal metallicity stars,
which means that the serious zero-point discrepancy found in old studies has been settled, 
This clearly indicates that applying the non-LTE corrections (and inclusion of Balmer 
line wings as background opacity) is indispensable for reliable Al abundance 
determinations of A-type stars from Al~{\sc i} 3944/3961 lines.

The resulting Al abundances of A-type stars are almost in proportion to [Fe/H] 
(tending to be overabundant in Am stars) with an approximate scaling relation of 
[Al/H]~$\sim 1.2$~[Fe/H]. This consequence is qualitatively consistent with 
the prediction of the diffusion theory (suggesting an Al excess in the photosphere 
of Am stars).\footnote{It should be remarked, however, that the situation is different 
for HgMn stars (chemically peculiar stars in the regime of late B-type stars), 
for which Al tends to be deficient as separately described in Appendix~A for 
the case of the AR~Aur system.} 

As a by-product of this study, the applicability of other Al lines (Al~{\sc i}~6696/6698,
Al~{\sc ii} 3900, and Al~{\sc ii} 4663) for Al abundance determination of A-type stars 
was also investigated. The Al~{\sc ii} 3900 line turned out to be unsuitable 
because it is badly blended with the strong Ti~{\sc ii} 3900 line. Yet, the 
Al~{\sc i}~6696/6698 lines (for late A-type stars; with a mild non-LTE correction 
of 0.1--0.3~dex) and Al~{\sc ii} 4663 line (for early A-type stars; almost free 
from the non-LTE effect) may be usable as Al abundance indicators, though  
limited to only sharp-lined stars because of their weakness.

\acknowledgements

This investigation has made use of the SIMBAD database, operated by CDS, 
Strasbourg, France, and the VALD database operated at Uppsala University,
the Institute of Astronomy RAS in Moscow, and the University of Vienna.

\clearpage
\appendix{Aluminium abundances of the AR Aur system}

In the author's previous non-LTE abundance determinations of CNO (covering 
up to $T_{\rm eff} \la 11000$~K; Takeda et al. 2018) and of Si (up to 
$T_{\rm eff} \la 14000$~K; Takeda 2022), which were based on the spectra 
obtained with HIDES and BOES like this study, it was possible to derive 
the abundances of HgMn stars (important group of chemically peculiar stars 
at 10000~K~$\la T_{\rm eff}$). 
Unfortunately, the program stars of this investigation could contain 
only A-type stars of $T_{\rm eff} \la 10000$~K (and HgMn stars are lacking), 
because the HIDES data of late B-type stars used in the past studies do not 
include the short-wavelength region covering the Al~{\sc i} 3944/3961 lines.    

However, a set of BOES spectra (of wide wavelength coverage) for AR~Aur,
a double-lined eclipsing binary comprising HgMn star (primary [P], B9V) and 
normal star (secondary [S], B9.5V), are available (Takeda et al. 2019). 
Therefore, as a supplementary analysis, non-LTE Al abundances of AR~Aur (P) 
and AR~Aur (S) were determined by applying the spectrum-fitting analysis 
to Al~{\sc i} 3944, Al~{\sc i} 3961, and Al~{\sc ii} 4663 lines as done 
in Sect~3.3, in order to check whether any difference exists in Al abundances 
between normal and HgMn stars. The spectra of both stars were disentangled 
as described in Sect.~2.2 of Takeda et al. (2019), and the same atmospheric 
parameters as well as model atmospheres as used in that paper were adopted. 

The results are summarized in Table~5, and the accomplished spectrum fit 
in each region is shown in Fig.~12, from which the following consequences 
can be drawn.
\begin{itemize}
\item
As seen from Fig.~12, the Al lines are very weak and hardly detectable in 
AR~Aur (P), which means that reliable abundance determination is not feasible. 
Still, $A$(Al)$\la 5.5$ (i.e., [Al/H]~$\la -1$) may be concluded from Table~5 
as the photospheric Al abundance for this HgMn star. 
\item
In contrast, all three Al lines are clearly observable in AR~Aur (S), and
a near-solar abundance of $A$(Al)~$\simeq 6.6$ ([Al/H]~$\simeq +0.1$) is 
consistently obtained for this normal B9.5V star.
\item
In summary, while the secondary star has quite normal abundances for both 
Fe and Al ([Fe/H]~$\simeq$~[Al/H]~$\simeq +0.1$), Al is considerably deficient
in the primary HgMn star ([Al/H]~$\la -1$) despite that it is Fe-rich ([Fe/H]=~+0.5).
\item
Two chemical abundance studies (including Al) on AR~Aur based on LTE are 
already published. Khokhlova et al. (1995) derived  [Al/H] = $-1.2$ (P) and 
$+0.5$ (S) from Al~{\sc i} 3944/3961 lines. Meanwhile, Folsom et al. (2010) 
concluded [Al/H] =  $-1.37$ (P) and $+0.30$ (S) (though the adopted 
spectral lines are not explicitly described). While their results for the 
primary (marked Al deficiency by $\la -1$~dex) are quite consistent with 
the consequence here, those for the secondary (moderate Al excess by several 
tenths dex) appears to be somewhat overestimated.  
\item 
Accordingly, unlike Am stars (for which Al-excess is accompanied with an 
overabundance of Fe), photospheric Al abundances in HgMn stars can be
considerably deficient even for the case of supersolar Fe. This means
that these two groups of chemically peculiar stars are markedly different
as long as the behavior of Al abundance is concerned.
\item
This observational fact is a reconfirmation of the results of Smith (1993),
who concluded from the analysis of UV lines based on IUE spectra
that Al is deficient (becoming increasingly underabundant at higher 
$T_{\rm eff}$) in essentially all HgMn stars. 
\end{itemize}

\setcounter{table}{4}
\begin{table}[H]
\scriptsize
\caption{Analysis results of AR~Aur primary (P) and secondary (S).}
\begin{center}
\begin{tabular}
{ccccc c@{ }c@{ }c c@{ }c@{ }c c@{ }c@{ }c}\hline \hline
Star & $T_{\rm eff}$ & $\log g$ & $\xi$ & [Fe/H] & 
$W^{\rm N}_{3944}$ & $A^{\rm N}_{3944}$ & $\Delta_{3944}$ & 
$W^{\rm N}_{3961}$ & $A^{\rm N}_{3961}$ & $\Delta_{3961}$ & 
$W^{\rm N}_{4663}$ & $A^{\rm N}_{4663}$ & $\Delta_{4663}$ \\ 
  &  (K) & (dex) & (km~s$^{-1}$) & (dex) & 
 (m\AA) & (dex) & (dex) &
 (m\AA) & (dex) & (dex) &
 (m\AA) & (dex) & (dex) \\
\hline
AR~Aur (P) & 10950 &  4.33  &  1.0  &  +0.47  &  (0.6) & (4.62) &  +0.23 &  (4.0) &(5.49)& +0.33 &(1.5) &(5.05)& +0.02 \\
AR~Aur (S) & 10350 &  4.28  &  1.6  &  +0.07  &  48.1  &  6.55  &  +0.30  &  42.8 & 6.58 & +0.42 & 24.7 & 6.66 &  +0.03\\
\hline
\end{tabular}
\end{center}
The parenthesized data derived for AR~Aur (P) are subject to large uncertainties 
(because the line is considerably weak and barely detectable) and thus unreliable.
\end{table}

\begin{figure}[H]
\centerline{\includegraphics[width=0.8\textwidth,clip=]{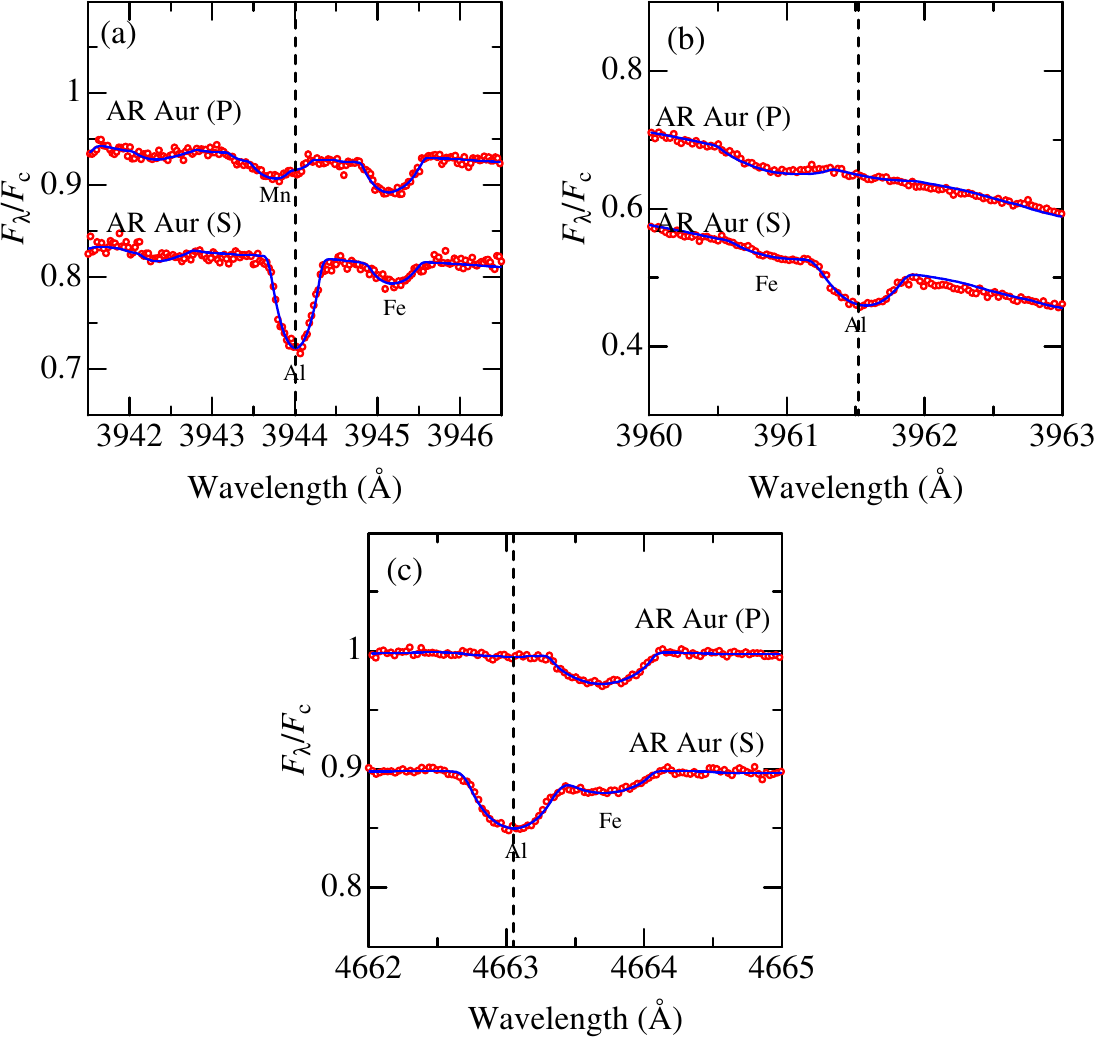}}
\caption{
Synthetic spectrum fitting analysis for Al abundance determinations
of AR~Aur (P) and AR~Aur (S) based on (a) Al~{\sc i} 3944,
(b) Al~{\sc i} 3961, and (c) Al~{\sc ii} 4663.
The wavelength positions of the relevant Al lines are shown by vertical dashed lines.
The scale of $F_{\lambda}/F_{\rm c}$ indicated in the left axis is 
for AR~Aur (P), wile the spectrum for AR~Aur (S) is shifted downward by 0.1.
Otherwise, the same as in Fig.~7.
}
\label{fig12}
\end{figure}

\end{document}